\documentclass[12pt,fleqn,psfig]{article}
\usepackage{amsmath, amssymb, graphicx}
\usepackage[hang,scriptsize,bf]{caption}
\usepackage{cite}

\usepackage{geometry}
\usepackage{color}
\usepackage[latin1]{inputenc}
\usepackage{mathrsfs}
\usepackage{amsfonts}
\usepackage[title]{appendix}

\begin{document}

 
\thispagestyle{empty}
\renewcommand{\thefootnote}{\fnsymbol{footnote}}

\begin{center} \noindent {\bf A new look at the Helmholtz equation:\\Lefschetz thimbles and the einbein action\footnote{MSC: Primary 35J05; Secondary 58K35. \\ \indent\indent Keywords: Helmholtz equation, Path Integral, Catastrophes, Monodromy, Picard-Lefschetz theory.}}

\end{center}

\bigskip\bigskip\bigskip

\bigskip\bigskip\bigskip

\begin{center}
Z.~Guralnik\footnote{guralnikz@leidos.com} \\
\it \small Leidos, Inc. 11951 Freedom Dr, Reston, VA, 20190
\end{center}
\bigskip

\bigskip\bigskip\bigskip\bigskip

\renewcommand{\thefootnote}{\arabic{footnote}}

\centerline{ \small Abstract}
\bigskip

\small 
Picard-Lefschetz theory is applied to solutions of the Helmholtz equation, formulated in terms of sums of integrals of a proper-time, or `einbein', wave function $\Psi(\Lambda) = \exp(i\mathbb S(\Lambda))$ along complex contours bounded by essential singularities of $\Psi$.  
There is a one to one map between steepest descent paths connecting  essential singularities and real or complex eigenrays. Residues of finite $\Lambda$ poles of $\mathbb S(\Lambda)$ are shown to vanish at spatial points corresponding to sources, provided that the pole bounds only one steepest descent path. If the sum  includes two such paths, with one beginning and the other ending at the same pole, points of vanishing residue are not sources, but are argued to be the locus on which caustic curves may have singularities such as cusp points.  
The map between $\mathbb S$ and the generating function in the Thom--Arnold classification of catastrophes is discussed.
Monodromies of the solution set with respect to complexified parameters defining the index of refraction, or spatial endpoints of Green's functions, are trivially determined from the singularities of $\mathbb S(\Lambda)$.  We construct a variant of a Laurent series expansion of $\mathbb S$ about a pole  at finite $\Lambda$. Expressions for the coefficients of each order in this expansion can often be given exactly.  Based on the Laurent series expansion, we propose a variation of a Pad\'e approximant for $\mathbb S$, with the intent of capturing additional poles and the associated cusp caustics which are not visible in the Laurent series expansion.

\newpage
\section{Introduction}


The wave propagation phenomena described in this article are studied by an approach having its origins in the Feynman path integral \cite{Feynman1,Feynman2}. 
Path integral solutions of the wave equation with a spatially varying index of refraction have been considered primarily in the context of ocean acoustics, although  the domain of interest potentially extends to many other fields such as optics or gravitational lensing.  Typically, the path integral is applied to a parabolic equation approximation to the Helmholtz equation, in which waves are assumed to be propagating nearly in one direction. Evaluation of the path integral in a quadratic order expansion about the real stationary phase paths, or eigenrays, has been used to great effect for short-wave length propagation problems in a stochastic ocean environment \cite{Dashen,Palmer}.
However caustics, as well as shadow zones into which no eigenrays penetrate, are not accounted for in this approximation. Numerical evaluation of the path integral solution to the parabolic equation\cite{Tappert1,Tappert2,Tappert3}, as well as variants to capture larger variations in propagation angle using Pad\'e approximants \cite{RGreene,Collins}, are adequate for computing fields at caustics and within shadow zones.  

In fact the path integral  is more than the basis for a numerical tool, containing an enormous amount of information about the solution set, shadow zone fields, caustics, and monodromies relating linearly independent solutions. The key to unlocking this information is the consideration of complexified path integration cycles. There are homology classes of equivalent cycles, having steepest descent representatives known as Lefschetz thimbles which pass through critical points. The Lefschetz thimbles and the dependence of the associated homology classes on variations of parameters of the problem, such as the index of refraction and spatial coordinates in the present context, is the subject of Picard-Lefschetz theory \cite{Vasiliev,Picard,Lefschetz1,Lefschetz2}.
This article considers these Lefschetz thimbles for the path integral representations of the Helmholtz equation, absent any approximations such as that leading to the parabolic equation.

The natural means to compute fields in a shadow zone where no real eigenrays exist involves deforming the integration path into the complex plane.
Although the integration over real paths is still viable in the shadow zone, the critical points of the action which dominate the result at short wavelength are complex.  These critical points are complex eigenrays, which have been considered without reference to path integrals in \cite{Keller,Keller2,KravtsovComplex,Wang,Chapman}.  
Both real and complex eigenrays lie on Lefschetz thimbles, on which the phase of the integrand or real part of the action is a constant.   The integral over real paths is equivalent to a sum of integrals over Lefschetz thimbles, which are more rapidly convergent  due to the lack of phase oscillations.  Lefschetz thimbles have been of considerable use in quantum field theory \cite{GG1}--\cite{Cristoforetti:2012su}, although they are somewhat difficult to construct in the case of functional integration. 

Fortunately there is an alternative solution of the Helmholtz equation involving integration cycles in just a single complex variable rather than an infinity of complex variables, specifically the complexification of the Schwinger proper time, denoted here by $\Lambda$.  The 
application of the Fock-Schwinger proper time formulalism \cite{Fock,Schwinger} to the Helmholtz equation is well known \cite{Botelho,Samelsohn,Fishman,Schlottmann}, albeit for an integration over the positive real axis.  This form of the solution was also considered even earlier in \cite{Holford1}, using different nomenclature, in which a number of the components of the results described here were presaged. 
The integrand is $\Psi = \exp(i{\mathbb S})$ where the action ${\mathbb S}(\Lambda)$ has critical points with a one to one map to eigenrays, both real and complex. 
The steepest descent paths in $\Lambda$ map to Lefschetz thimbles passing through eigenrays in the path integral.  We shall therefor also refer to steepest descent paths in $\Lambda$ as Lefschetz thimbles, even though that term is usually reserved for integration cycles involving many complex variables.

The proper time and eigenray approaches to the Helmholtz equation are in fact distinguished by a choice in the order of integration in a ``parent" path integral, discussed in section \ref{sec2}, containing phase space paths $\vec X(\tau),\vec P(\tau)$ and an einbein field $\Lambda(\tau)$. The latter is Lagrange multiplier enforcing re-parameterization invariance with respect to the path parameter $\tau$.  Reparameterization invariance implies a vanishing Hamiltonian, which is equivalent to the eikonal equation of ray theory.  If one instead integrates over phase space paths $\vec X(\tau),\vec P(\tau)$ first, fixing a gauge in which $\Lambda(\tau)$ is independent of $\tau$, one obtains the Schwinger proper time formulation, in which ${\mathbb S}(\Lambda)$ will be referred to as the einbein action.
The phenomena of sources, caustics, shadow zones, complex rays, ray generation under perturbation, as well as monodromies relating linearly independent solutions, have a unifying description  in terms of Lefschetz thimbles, either in the  complexified phase space or in the complex $\Lambda$ plane.

The proper time integral representation for a solution of the Helmholtz equation is 
\begin{align}\label{greensprop}
\phi(\vec x) = \frac{i}{k_0}\int_\Gamma d\Lambda \Psi(\Lambda,\vec x,\vec x')\, ,
\end{align}
where $\Psi$ satisfies a Schr\"oedinger equation,
\begin{align}\label{Schrod}
\frac{i}{k_0} \frac{\partial}{\partial\Lambda}\Psi+ \left(\frac{1}{k_0^2}\vec{\nabla}_x^2 +  n(\vec x)^2\right)\Psi =0\, .
\end{align}
The Helmholtz operator $\frac{1}{k_0^2}\vec{\nabla}^2 +  n(\vec x)^2$ is analogous to a Hamiltonian, with $1/k_0$ playing the role of Planck's constant and $\Lambda$ the role of time. Cases in which $\Psi(\Lambda)$ can be given exactly were first described in \cite{Holford1}, and are reviewed in sections \ref{const}--\ref{schan}, in which they are derived using a path integral.
Equation \eqref{Schrod} differs from the Schr\"oedinger equation of quantum mechanics in that the Hamiltonian is not bounded below and, more importantly, the only solutions of \eqref{Schrod} of interest are those for which there are integration cycles $\Gamma$ yielding non-trivial solutions of the Helmholtz equation.  Such solutions turn out to have have essential singularities at finite values of $\Lambda$. 
From \eqref{Schrod} it follows that 
\begin{align}\label{surff}
\left(\vec{\nabla}_x^2 +  k_0^2 n(\vec x)^2\right) \int_\Gamma d\Lambda \Psi =  -ik_0 \Psi \big |^{\Gamma_+}_{\Gamma_-}
\end{align}
for any contour $\Gamma$  bounded by $\Gamma_{-}$ and $\Gamma_+$. Although $\Gamma$ is ordinarily taken to be the positive real axis, yielding a Green's function of the Helmholtz equation satisfying the Sommerfeld radiation boundary conditions,  there are non-trivial homology classes, each containing a set of  complex contours equivalent via Cauchy's theorem, which correspond to different solutions of the Helmholtz equation. 
In addition to solutions of the Helmholtz equation obtained from closed contours encircling singularities, one can also consider $\Gamma$  bounded by essential singularities, approached such that $\Psi(\Lambda)$ vanishes along with the right hand side of \eqref{surff}. 

In all the soluble examples, the analytic structure of the einbein action ${\mathbb S}$ has the form
\begin{align}\label{poleslogs}
{\mathbb S} \equiv -i\ln \Psi(\Lambda)= simple\,\, poles + logs + entire.
\end{align}
It will be shown in section \ref{laurent} that coefficients of Laurent series expansions of the einbein action about the poles have a remarkably simple analytic dependence on the wave-number, valid in both the small and large wave-number limits. 
A variation of a Pad\'e approximant for $\mathbb S(\Lambda)$ can be used to capture the existence of other poles.  This approximant is expected to be very powerful, for $\mathbb S(\Lambda)$ having no singularities other than poles or logarithms.
The poles and logarithms are closely related to each other via \eqref{Schrod}, in a way which is built into the approximant. 
A property of the exactly soluble cases is that there are no singularities of $\mathbb S$ besides poles and logarithms, and there are also no caustics of higher order than cusps. 
The more general case will certainly contain higher order caustics, and we can not exclude the possibility of other singularities, perhaps essential, of $\mathbb S$. In the absence of these other singularities however, the Pad\'e approximant can be expected to be quite powerful.

An integral of $\Psi(\Lambda)$ over real $\Lambda$ is highly oscillatory and marginally convergent, but may be deformed into an  equivalent and much more rapidly convergent integration over a sum of complex contours $\Gamma_i$ connecting the poles of $\mathbb S$.  
At large $k_0$, $\Psi$ behaves as  
\begin{align}\label{barS}
\Psi(\Lambda) = f(\Lambda)\exp(ik_0\bar S(\Lambda)),
\end{align} 
where $\bar S$ is meromorphic in $\Lambda$,  logarithmic terms in $\mathbb S$ having been absorbed into the definition of $f(\Lambda)$. We shall often refer to $\bar S$, which is the leading term in a large $k_0$ expansion of $\mathbb S/k_0$, as the einbein action. The integration contours $\Gamma_i$ are Lefschetz thimbles on which ${\rm Re}(\bar S)$ is constant, passing through critical points where $\frac{d{\bar S}}{d\Lambda}=0$.
The poles and critical points of $\bar S$ are  related by the Riemann--Hurwitz formula.
There is a one--to--one map between the critical points of $\bar S$ and rays, both real and complex.   At leading order in the $1/k_0$ expansion,  \eqref{Schrod} implies that $\bar S$ evaluated at a critical point is a solution of the eikonal equation; 
\begin{align}
(\vec\nabla_x{\bar S})^2 + n(\vec x)^2 =  - \frac{d{\bar S}}{d\Lambda} =0\, ,
\end{align}  
hence the map between critical points of $\bar S(\Lambda)$ and eigenrays.  
However the means to compute $\bar S( \Lambda)$ do not involve ray tracing. The summation over rays is supplanted here with a summation over Lefschetz thimbles $\Gamma_i$ in the complex $\Lambda$ plane. 

A general solution of the Helmholtz equation may be written as 
\begin{align}
\phi&= \frac{i}{k_0}\sum_i c_i  \int_{\Gamma_i} d\Lambda \Psi(\Lambda)
\end{align}
including all inequivalent contours $\Gamma_i$ with arbitrary complex coefficients $c_i$. The Lefschetz thimbles change discontinuously upon crossing a caustic, at which critical points in the complex $\Lambda$ plane coalesce. For instance, upon crossing a fold caustic from an illuminated region to a shadow zone, two real critical points contributing to the solution merge and then split into two complex critical points,  only one of which contributes to the Green's function in a shadow zone.  




There is always at least one simple pole of $\mathbb S \equiv -i \ln \Psi$, having a universal residue independent of the index of refraction, as shown in section \ref{laurent}. Since \eqref{Schrod} is invariant under a constant shift of $\Lambda$, this pole is chosen to lie at $\Lambda=0$, coinciding with the location of the pole found in a path integral derivation of $\Psi$.  This residue vanishes along some spatial locus, e.g. $\vec x-\vec x'=0$, in which case a contour ending at $\Lambda=0$ has a non-zero endpoint contribution in \eqref{surff}, 
\begin{align}
\lim_{\Lambda\rightarrow 0}\Psi(\Lambda)=
\delta(\vec x-\vec x')\, ,
\end{align}
where the limit is taken so as to approach the essential singularity of $\Psi$, or pole of $\mathbb S$, in a convergent direction.  Thus a contour ending at $\Lambda=0$ yields a delta function source and a Green's function of the Helmholtz equation. 
The Green's function which satisfies the radiation condition is given by the integral over the positive real axis, which is equivalent to a particular sum over Lefschetz thimbles $\Gamma_i$, including one terminating at $\Lambda=0$.

In addition to the universal pole of $\mathbb S$ at $\Lambda=0$, is will be shown that there may be other simple poles at finite $\Lambda$.
The residues of the finite $\Lambda$ poles  are  functions of $\vec x$ and $\vec x'$ which are very similar to the residue of the universal pole at $\Lambda=0$, vanishing at points  which we refer to as `ghost sources'.   Each ghost pole bounds a pair of oppositely oriented contours, acting as the sum of a source and a sink such that there is no additional inhomogeneous term in the Helmholtz equation. 
One can not move these pairs of contours away from the ghost pole if the parts ending and beginning at the pole lie on different Riemann sheets of $\Psi$.  
Even if one can deform a contour away from a ghost pole, the most useful representative of a given homology class is a sum over Lefschetz thimbles, which may include pairs with endpoints at the same pole. 
Although lacking a direct physical interpretation,  ghost poles and ghost sources will be shown to be intimately related to both monodromies and cusp caustics.


New poles, and therefore new rays, may be generated either by deformations of $n(\vec x)$ or of a source $J(\vec x)$.  Via arguments given in section \ref{laurent}, new poles can only enter from $\Lambda=\infty$ under deformations of $n(\vec x)$.   At finite $\Lambda$, new poles can appear by the splitting of existing poles due to smearing of a delta function source, as shown in section \ref{cusp}.  Pole splitting is invariably accompanied by the formation of a cusp caustic.  
Under very general conditions, pairs of nearby poles lead to cusp caustics. In examples considered in section \ref{cusp} and \ref{class}, the cusp points lie along curves corresponding to a ghost source.  We conjecture that this is true in general; cusp caustics always coincide with ghost sources.


In section \ref{unif},  a uniform asymptotic approximation for a smooth caustic is derived using the einbein action formulation.  
Uniform asymptotic approximations are described in \cite{Pearcey,Chester,Kravtsov1,Kravtsov2,Ludwig,Berry,Berry2,Berry3} and a particularly lucid review can be found in \cite{Holford2}.  
Uniform asymptotic approximations are essentially fixed by the local geometry of caustics and nearby real eigenrays \cite{Ludwig}, such as the difference between the curvatures of a fold caustic and that of the intersecting rays \cite{Holford2,Spofford}.  Although based on real rays, the contribution of complex rays is implicit in the result.  However there are conditions, particularly at larger wavelengths, under which there is reason to explicitly consider complex rays or the corresponding complex critical points of the einbein action.  Fourier transforming the Green's function of the Helmholtz equation with respect to $k_0$ yields the signal due to a pulse, temporally separating contributions due to rays. 
The arrival time is proportional to the real part of the einbein action on the associated Lefschetz thimble, while  temporal smearing is related to the imaginary part of the action at the critical point.  Even in illuminated regions, there may be arrivals due to complex eigenrays which are often neglected, as has been emphasized in \cite{Chapman}. The existence and importance of complex saddle points is particularly transparent in the einbein formulation.

As observed in \cite{Maslov,Kravtsov,Duistermaat,Berry4,KravtsovOrlov2,KravtsovOrlov1}, the uniform asymptotic approximation to the field in the neighborhood of a caustic is very closely related to the classification of catastrophes due to Thom and Arnold \cite{Thom,Arnold}.   In principle, caustics corresponding to catastrophes of the $A_N$ type should occur  naturally in the einbein formulation;  there is a locally defined map $\Lambda \rightarrow \lambda$ relating the einbein action $\bar S(\Lambda)$ to the generating polynomial $P(\lambda)$ defining the catastrophe. The relation between the einbein description and Thom--Arnold classification of non-smooth caustics is discussed in section \ref{class}. 
As discussed in section \ref{unif}, the map is non-singular at smooth caustics. However the map is singular at the intersection of the smooth caustic curves with ghost sources, which coincides with cusp points.  In fact a special case of the map at a cusp was given in \cite{Holford1}, using methods described in \cite{Bleistein}. 

Generating functions of the $D_N$ and $E_6,E_7,E_8$ catastrophes are polynomial in two variables.  An integral representation with two einbeins is possible for scattering problems in which two Green's function are coupled by a scattering kernel, but does not occur naturally for an analytic index of refraction.  However, scattering problems will not be considered here.   There are also more exotic caustics having a generating function which is a polynomial in more than two-variables, for which we shall offer no einbein interpretation.



Linearly independent solutions of the Helmholtz equation, as well as different eigenrays, may be  related by analytic continuation of 
parameters defining the index of refraction $n^2(\vec X)$, or the spatial coordinates, around closed loops about branch points in the complex plane.  In section \ref{monodromies}, we show how  these relations, or monodromies, are trivially determined from the singularities of $\mathbb S(\Lambda)$.  
The integration cycles in $\Lambda$ are not invariant under closed loop variations of the parameters.   
Convergence of  $\int d\Lambda\Psi$ requires that integration contours  only approach essential singularities of $\Psi$ within certain angular domains.  These domains change as the argument of various parameters are varied from $0$ to $2\pi$,  inducing non-zero winding numbers around the finite poles, as well as jumps between convergence domains of the pole at infinity.  Unlike the finite poles, which are simple, the pole at infinity may be higher order.

Complex integration cycles are also familiar in  quantum field theory.  Path integrals over suitable multi-dimensional complex curves  are solutions of the Schwinger Dyson equations, Schwinger action principle and Ward identies\cite{GG1,GG2,WIT,WIT2,Ferrante:2013hg}.  
These integration cycles have been of much interest in the context of resurgence theory \cite{Dunne:2015eaa,Behtash:2015loa} and as potential solutions of the `sign problem' \cite{Pehlevan:2007eq,Guralnik:2009pk,Cristoforetti:2012su}. 
The path integrals in this case are bounded by poles of the action at infinity in the space of complex quantum fields.
Finite poles, which are generic in the wave propagation problem, are not encountered in standard quantum field theory applications of Picard-Lefschetz theory, in which the action is an entire function.

\section{Remarks on path integral solutions of the Helmholz equation}\label{sec2}

There exist a number of path integral representations of the Green's function of the Helmholtz equation,  
\begin{align}
\left(\vec\nabla_x^2 + k_0^2 n(\bf x)^2\right)G(\vec x,\vec x') = \delta^D(\vec x - \vec x')\, . 
\end{align}
The most familiar path integral representation, especially in the context of ocean acoustics, is based on the parabolic approximation, in which one picks a particular `forward' direction $x$. Writing
\begin{align}
G= e^{ik_0 x} \tilde G
\end{align}
the Helmholtz equation is approximated by,
\begin{align}\label{parabolic}
\left(2ik_0\partial_x + \vec\nabla_y^2 + k_0^2(n(\vec x)^2 - 1)\right)\tilde G = \delta(\vec x -\vec x') 
\end{align}
where $\vec x= (x,\vec y_\perp)$,
and the term $\partial_x^2 \tilde G$ has been dropped under the sometimes false assumption that the $\tilde G$ is very slowly varying with $x$ compared to $\exp(ik_0 x)$.   Since \eqref{parabolic} has the form of a Shr\"oedinger equation, with $x$ analogous to time, a path integral representation follows naturally \cite{Dashen,Palmer},
\begin{align}
\tilde G &= <\vec y'| e^{ik_0\left(\frac{\vec\nabla_y^2}{2k_0^2} + \frac{1}{2}(n^2-1)\right)} |\vec y> \\
&= \int D\vec Y_\perp(X) D\vec P_\perp(X) e^{ik_0\int_{x'}^{x} dX\left(\vec P_\perp\cdot \frac{ dY_\perp}{dX} - \frac{1}{2}\left( \vec P_\perp^2 - (n(X,\vec Y_\perp)^2-1) \right)\right)}\, ,
\end{align}
where the path $\vec Y_\perp(X)$ is bounded by $\vec y_\perp$ and ${\vec y_\perp}^{\,'}$ at the points $X=x$ and $X=x'$.  The  canonical momenta  $\vec P_\perp(X)$, related to the angle of the wavefront, are unconstrained at the endpoints.
This representation has proven particularly useful for long range propagation problems in the presence of stochastic fluctuations of the index of refraction $n(X,\vec Y_\perp)$,  in which the path integrals are evaluated in an expansion about deterministic ray paths \cite{Dashen}.  A path integral formulation for the Helmholtz equation, in the absence of any one-way propagation approximation,  also exists and is reviewed below.

In the Fock-Schwinger proper time formulation, the 
Green's function of the Helmholtz equation, or matrix elements of the inverse Helmholtz operator, is written as
   \begin{align}\label{FS}
  G(\vec x,\vec x') =  <\vec x|\left({\vec{\nabla}_{\bf X}}^2 + k_0^2 n(\vec{\bf X})^2\right)^{-1} |\vec x'>=\frac{i}{k_0}\int_0^\infty d\Lambda\,  <\vec x|e^{ik_0\Lambda\left(\frac{1}{k_0^2}\vec{\nabla}_{\bf X}^2 +  n(\vec {\bf X})^2\right)}|\vec x'>\, . 
   \end{align} 
The path integral representation of \eqref{FS}, obtained by standard methods \cite{Feynman2},  is given by
\begin{align} 
G = \int_0^\infty &d\Lambda \int {\cal D} \vec X(\tau) {\cal D}\vec P(\tau) e^{ik_0\int_0^1 d\tau{\cal L}} \nonumber\\ 
{\cal L} &\equiv 
\vec P \cdot \frac{d\vec X}{d\tau} - \Lambda(\vec P^2 - n(\vec X)^2)
\, ,\label{pthorig}
\end{align} 
where the paths $\vec X(\tau)$ are bounded by $\vec X(0)=\vec x', \vec X(1)=\vec x$. For certain $\Lambda$,  corresponding to the ghost poles to be described later, the Euler Lagrange equations associated with the Lagrangian ${\cal L}$ have no solution for generic Dirichlet boundary conditions on $\vec X$.  The particular boundary conditions admitting a solution are the locus of vanishing residue of the ghost poles, or ghost sources. 

The parameterization of the paths $\vec X(\tau), \vec P(\tau)$ has no physical meaning. One can make any redefinition $\tau\rightarrow\tau'(\tau)$, together with a redefinition of $\Lambda$ such that it becomes a field dependent on $\tau$, transforming so that $\Lambda(\tau)d\tau$ is invariant.
One choice, $d\tau = n(\vec x)\sqrt{d\vec x^2}$, leads to an interpretation of $\tau$ as physical time.  Another choice, $X(\tau)= X(0) + (X(1)-X(0))\tau$, is related to the parabolic approximation, and is singular for paths that are not monotonic in $X$. The path integral \eqref{pthorig} is a `gauge fixed' version of a more general formulation in which $\Lambda$ is a field dependent on $\tau$,
\begin{align}\label{FullPI}
G = &\int  {\cal D} \vec X(\tau) {\cal D}\vec P(\tau) {\cal D}\Lambda(\tau)\,\, e^{ik_0S[\vec X,\vec P, \Lambda]} \\ 
&S \equiv \int_{\tau=0}^{\tau=1} d\tau \left(\vec P \cdot \frac{d\vec X}{d\tau} - {\Lambda}(\vec P^2 - n(\vec X)^2)\right)\, . \label{action}
\end{align}
This form involves an infinitely redundant integration over equivalent paths, hence the need for gauge fixing which may be carried out in a number of ways using the Fadeev--Popov procedure \cite{FaddeevPopov}, the technicalities of which are beyond the scope of this discussion.  The gauge $d\Lambda/d\tau =0$ yields the Fock-Schwinger-Feynman formulation.   
Aside from a metric signature and the dependence of $n^2$ on $\vec X$, \eqref{FullPI} resembles the path integral for a massive particle. In the particle physics nomenclature, $\Lambda$ would be referred to as an `einbein', and we shall do the same here.  

As an aside, note that Fourier transforming \eqref{FullPI}  with respect to an endpoint variable $ x_i$ is equivalent to another path integral in which the endpoint $P_i(\tau=1)$ is fixed, whereas $X_i(\tau=1)$ is un-constrained.  The saddle points in this case yield geometric optics solutions in a hybrid position--wavenumber space $(P_i, X_{j\ne i})$.  Such hybrid solutions are used in Maslov's approach  \cite{Maslov1,Maslov2,Maslov3} to computing the field in the neighborhood of a caustic, at which geometric optics in position space is singular. 

The  critical points of the action \eqref{action}, or eigenrays, are solutions of the equations of motion,
\begin{align}
\frac{\delta S}{\delta X} &= 0  \rightarrow \,\, -\frac{d\vec p}{d\tau} + {\Lambda}\vec\nabla\left( n(\vec X)^2 \right) = 0 \label{eqns1}\\ 
\frac{\delta S}{\delta P} &= 0 \rightarrow \,\, \frac{d\vec X}{d\tau} -2 \Lambda \vec P =0 \label{eqns2} \\
\frac{\delta S}{\delta \Lambda} &= 0 \rightarrow \,\, \rightarrow \vec P^2 - n(\vec X)^2 =0\, . \label{einbec}
\end{align}
Choosing $\Lambda =1/2$  leads to a standard first order form which is often used in numerical ray tracing. Note that $\Lambda$ plays the role of a Lagrange multiplier, enforcing the vanishing of the Hamiltonian, 
\begin{align} \label{HamVan}
H(\vec X,\vec P) \equiv \vec P^2 - n(\vec X)^2=0\, ,
\end{align} 
due to invariance under re-definitions of $\tau$. 
The eikonal, or Hamilton-Jacobi, equation for the action $S=S_{cl}$  satisfying \eqref{eqns1}--\eqref{einbec} is
\begin{align}
H(\vec x, \vec\nabla S_{cl})=(\vec\nabla_x S_{cl})^2 - n(\vec x)^2 = 0\, ,
\end{align}
where the gradient is with respect to the endpoint value  $\vec x =X(\tau=1)$, keeping the source location $\vec x'$ fixed. 
 
The vanishing Hamiltonian can be interpreted as the reason for the existence of caustics.  For a given initial point $\vec X(0) = \vec x'$, the canonical momentum or velocity $\frac{d\vec X}{d\tau} = \Lambda\vec P$ is constrained by $H=0$.  Thus for a non-constant $n(X)$, or non-zero acceleration, certain endpoints $\vec x$  will be inaccessible to real solutions.  These regions are the shadow zone, separated from the illuminated zone by a caustic.  Since $H=0$ follows from stationarity with respect to the einbein \eqref{einbec}, the presence of shadow zones and caustics is not manifest in a path integral formulation until after the einbein integration has been carried out. 

In the language of the wave equation with a position dependent phase velocity $c(\vec X) = \omega/(k_0n(\vec x))$, eigenrays are local minima of the travel time between paths connecting $\vec x$ and $\vec x'$.
However the travel time has no real extremum in a shadow zone.  Clearly, there is still a real solution to the minimum travel time problem, however it lies at boundaries of the domain of path integration.  For example, such paths could include a component propagating along a bounding surface.  These paths do not satisfy the equations of motion and are not critical points 
controlling the large $k_0$ behavior of the path integral.  
The critical points in the shadow zone are complex eigenrays.
The path integration may be deformed into an integral over complex cycles passing through critical points while keeping the phase of the integrand ${\rm Re}(S[\vec X(\tau),\vec P(\tau),\Lambda(\tau)])$ constant. These cycles, or Lefschetz thimbles, are  infinite dimensional versions of steepest descent paths \cite{WIT,WIT2} and are generally complex, even if the associated eigenray is real.  The number and topology of  Lefschetz thimbles changes as one crosses a caustic,  at which eigenrays coalesce. 

No effort will be made to explicitly construct Lefschetz thimbles in the complexified phase space.  
We shall instead consider their far simpler analogue in the complex  proper time plane.
Carrying out the functional integral over $\vec X$ and $\vec P$ before $\Lambda$ in \eqref{FullPI}, and fixing the 
$d\Lambda/d\tau=0$ gauge, gives the Schwinger proper time integral representation of the Green's function,
\begin{align}
G= \int d\Lambda e^{i{\mathbb S}(\Lambda)}
\end{align}
In cases in which $\mathbb S$ can be computed exactly, it is the sum of a meromorphic term and a logarithmic term which is subleading at large $k_0$,
\begin{align}
{\mathbb S} = k_0 \bar S + \ln(f).
\end{align}
Lefschetz thimbles can be defined with respect to $\bar S(\Lambda)$, whose critical points, at which $d\bar S/d\Lambda=0$, have a one to one map to eigenrays.  
The term einbein action will often be used in reference to $\bar S$ rather than the full action ${\mathbb S}$.

The path integral \eqref{pthorig} is Gaussian in $\vec P$, and also in $\vec   X$ provided that the index of refraction is quadratic,
\begin{align}
n(\vec X)^2 = n_0^2 + A_i X_i + B_{ij}X_i X_j\, .
\end{align}
In this case $\mathbb S(\Lambda)$ can be computed exactly, either via a path integral or methods described in \cite{Holford1}.
The exactly soluble class already includes non-trivial phenomena, including fold caustics, cusp caustics, eigenray generation and non-trivial monodromies.  The results for some such cases are shown below, in order to explicitly illustrate points made in the introduction.

\section{Constant $n^2$}\label{const}

Despite its apparent triviality, the case of constant index of refraction, $n(\vec X)=n_0$, is useful to demonstrate basic properties of the einbein action.  
Path integration over $\vec P(\tau)$ in \eqref{pthorig} yields
\begin{align}\label{PIconst}
G(\vec x',\vec x) &= \frac{i}{k_0}\int_0^\infty d\Lambda \int {\cal D} \vec X(\tau)  \,e^{ik_0{\cal L}[\vec X(\tau),\Lambda]}\\
&{\cal L}[\vec X(\tau),\Lambda]=\int_0^1 d\tau\left(  \frac{1}{4\Lambda}\left( \frac{d\vec X}{d\tau} \right)^2   + \Lambda n_0^2\right)\, ,  
\end{align}
where the integrated paths run between $\vec x'$ and $\vec x$. 
The Gaussian path integration over $\vec X(\tau)$ can also be carried out exactly, using methods described in \cite{Feynman1,Feynman2,Murayama}, giving
\begin{align}\label{constsoln}
G= & \frac{i}{k_0}\int_0^\infty d\Lambda\Psi = \frac{i}{k_0}\int_0^\infty d\Lambda f(\Lambda) e^{i k_0 \bf\bar  S}\nonumber\\
&f(\Lambda) \equiv \left(\frac{k_0}{4\pi i \Lambda}\right)^{D/2} \\ &\bar S={\cal L}[\vec X_c(\tau),\Lambda]\, ,
\end{align}
where $D$ is the number of spatial dimensions and the einbein action ${\bf\bar S}$ is the value of ${\cal L}$ at the critical point $\vec X(\tau)=\vec X_{c}(\tau)$, at which 
\begin{align}\label{actfu}
\frac{\delta {\cal L}}{\delta \vec X(\tau)} = -\frac{1}{2\Lambda}\frac{d^2\vec X}{d\tau^2} = 0\, .
\end{align}
The solution of \eqref{actfu} satisfying the boundary condition $\vec X(0)=\vec x',\,\vec X(1)=\vec x$,  is
\begin{align}
\vec X_c(\tau) = \vec{x'} + (\vec x-\vec x')\tau\, , 
\end{align}
so that
\begin{align}
{\bf\bar S} = {\cal L}[\vec X_c]=\frac{1}{4\Lambda}(\vec x-\vec x')^2 + \Lambda n^2\, . \label{constS}
\end{align}

\begin{figure}[!h]
	\center{
		\includegraphics[width= 200pt]{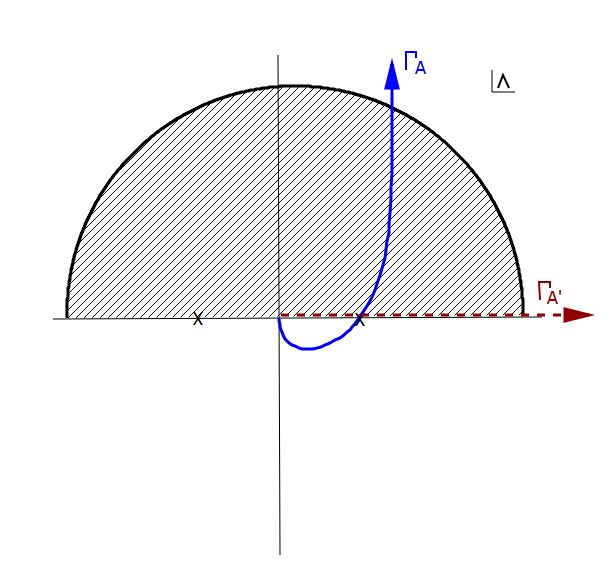}
		\caption{A steepest descent path or Lefschetz thimble $\Gamma_A$  in the complex $\Lambda$ plane for a constant index of refraction, on which ${\rm Re}(\bar S)$ is a constant and ${\rm Im}(\bar S)$ is a minimum at a critical point. The two critical points of $\bar S(\Lambda)$ on the real axis are marked by $X$.  Integration of $\Psi =f(\Lambda)\exp^{ik_0\bar S(\Lambda)}$ on $\Gamma_A$ is equivalent to integration along the positive real axis $\Gamma_{A'}$.  However integration along $\Gamma_{A'}$ is marginally convergent, whereas that along $\Gamma_A$ is rapidly convergent, dominated by the neighborhood of the critical point at positive real $\Lambda$.  The result is the Greens function satisfying the retarded or Sommerfeld-radiation boundary condition. Another steepest descent contour is given by the reflection of $\Gamma_A$ about the imaginary axis, passing through the other critical point and corresponding to an advanced boundary condition.}
		\label{fig:ConstIndex}
	}
\end{figure}

Given \eqref{constsoln} and \eqref{constS}
one finds
\begin{align}\label{demo}
\left(\vec\nabla_x^2 + k_0^2n^2\right)G =  \int_\Gamma d\Lambda \frac{\partial}{\partial\Lambda}\left[ \left(\frac{k_0}{4\pi i \Lambda}\right)^{D/2} e^{ i k_0 \left( \frac{1}{4\Lambda}(\vec x-\vec x')^2 + \Lambda n^2\right) } \right] \, ,
\end{align}
which is just the integrated Schr\"oedinger equation \eqref{surff}.
The right hand side of \eqref{demo} is $\delta(\vec x-\vec x')$, provided $\Gamma$ is a contour connecting the poles of $\bar S$ at $\Lambda=0$ and $\Lambda=\infty$.  
Convergence requires that the pole at $\Lambda=0$ is approached from below, within the wedge $-\pi<arg(\Lambda)<0$, while the pole at infinity is approached within the wedge $\pi>arg(\Lambda)>0$.  A  Lefshetz thimble equivalent to the positive real axis integration is illustrated in figure \ref{fig:ConstIndex}. This rendering is not exact, and is intended simply to show the approach to the poles.
The right hand side of \eqref{demo} vanishes so long as the residue of the pole at $\Lambda=0$ is non-zero, or $\vec x\ne \vec x'$.  To see that a delta fuction arises, let us move the $\Lambda=0$ endpoint of integration an infinitesimal amount to $\Lambda = -i\epsilon$.  
Then \eqref{demo} becomes
\begin{align}\label{deltfn}
\left(\vec\nabla_x^2 + k_0^2n^2\right)G =   
\left(
	\frac{k_0}{4\pi \epsilon}
\right)^{D/2} 
e^{   
	-\frac{k_0}{4\epsilon}(\vec x-\vec x')^2
}
\end{align}
which is a delta function in the $\epsilon\rightarrow 0$ limit.

For a position dependent index of refraction $n(\vec X)$,  we shall see that there are  generically additional finite poles of $\bar S$, or a higher order pole at infinity.   However the pole at $\Lambda=0$ has a universal form for a delta function source, for reasons to be made clear later, such that the leading term in a Laurent series about $\Lambda=0$ is always
\begin{align}
{\bf\bar S} = \frac{1}{4\Lambda}(\vec x-\vec x')^2 + \cdots \,\, .
\end{align}
Expressing the Greens function obtained by integrating over positive real $\Lambda$ as the sum of an integral over a number of Lefschetz thimbles connecting essential singularities, only the component with an endpoint at $\Lambda=0$ gives rise to the  delta function source in the Helmholtz--Greens equation.  When there are other finite $\Lambda$ poles,  the sum over thimbles equivalent to the positive real axis integration is such that these poles yield no additional inhomogeneous terms.  
Examples of such `ghost sources' are given later.


\section{Linear $n^2$}\label{sec3}

A two dimensional example of an index of refraction giving rise to a fold caustic is
\begin{align}\label{toy}
n^2(X,Z) = n_0^2 - aZ\, .
\end{align}
The Fock-Schwinger-Feynman proper time representation of the Greens function satisfying the radiation boundary condition is 
\begin{align}\label{PropX}
G(\vec x',\vec x)= &\frac{i}{k_0}\int_0^\infty d\Lambda \int {\cal D} \vec X(\tau)  \,e^{ik_0 {\cal L}[\vec X(\tau),\Lambda]}\, , \\ 
&{\cal L} = \int_0^1 \left(  \frac{1}{4\Lambda}\left( \frac{d  X}{d\tau} \right)^2  +   \frac{1}{4 \Lambda}\left( \frac{d  Z}{d\tau} \right)^2   + \Lambda (n_0^2-aZ) \right) d\tau \label{Om2}\, .
\end{align}
The Gaussian path integration in $\vec X$ can again be carried out exactly.  To this end, one solves the Euler Lagrange equations, $\frac{\delta {\cal L}}{\delta X(\tau)}=0$  or,
\begin{align}
\frac{1}{2\Lambda}\frac{d^2  X}{d\tau^2}  &=0 \\
\frac{1}{2\Lambda}\frac{d^2  Z}{d\tau^2} &+ a\Lambda  =0 \, .
\end{align}
For simplicity, consider the initial condition $ x' = z' =0$, in which case 
the solution is 
\begin{align}\label{toysoln}
X_c &=x\tau\, , \nonumber \\
Z_c &=   (z+\Lambda^2 a)\tau- \Lambda^2 a \tau^2 \, .
\end{align}
These trajectories are not to be confused with eigenrays.  There is no vanishing Hamiltonian constraint, having not carried out the integration over the einbein, hence there are real solutions for all $(x,z)$  with no indication yet of the presence of a shadow zone or a caustic.   Inserting the solution \eqref{toysoln} into \eqref{Om2} yields the einbein action 
\begin{align}\label{einbein_action}
{\bf \bar{\cal S}}(\Lambda,\vec x'=0,\vec x) = {\cal L}[\vec X_c(\tau),\Lambda]=\frac{1}{4\Lambda}\left(x^2 + z^2\right) + \Lambda\left( n_0^2 - \frac{a}{2}z\right) - \frac{1}{12}a^2\Lambda^3\, .
\end{align}

\begin{figure}[!h]
	\center{
		\includegraphics[width= 200pt]{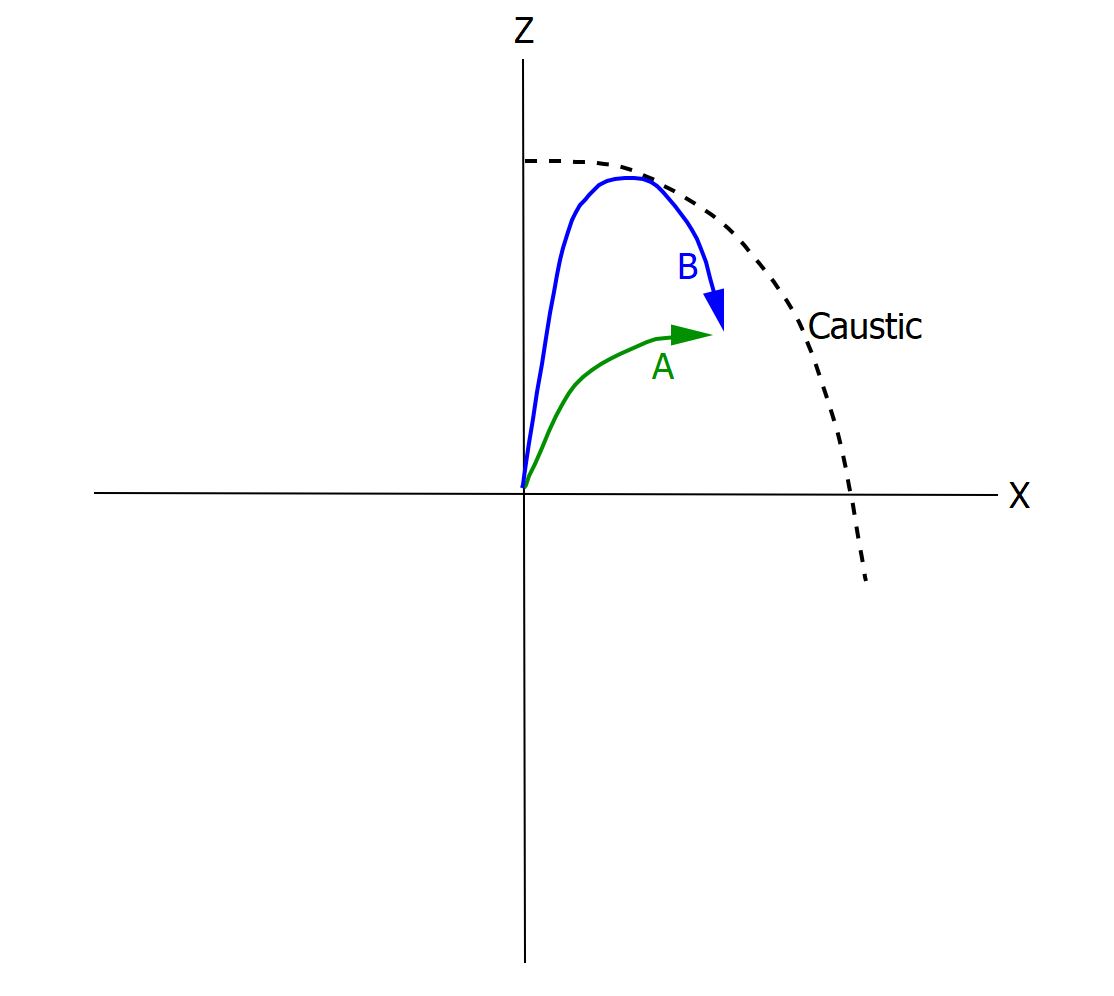}
		\caption{Two eigenrays in the illuminated zone ending at the same point $\vec x$, for the index of refraction $n(x,z)^2=n_0^2-Az$ and a source at the origin.}
		\label{fig:TwoRays}
	}
\end{figure}

The  path integral over $\vec X(\tau)$ yields
\begin{align}\label{LamEff}
G(\vec x'=0,\vec x) = \int_0^{\infty} d\Lambda \left( \frac{1}{4\pi \Lambda}\right) e^{ ik_0 \bar S(\Lambda,\vec x'=0,\vec x)} \, .
\end{align}
The integral over positive real $\Lambda$ is equivalent to a sum of  integrals over Lefschetz thimbles passing through critical points of $\bar S$.  The critical points satisfy 
\begin{align}
\frac{d {\bf \bar{\cal S}}}{d\Lambda} = -\frac{1}{4\Lambda^2}(x^2 + z^2) + (n_0^2-\frac{a}{2}z) - \frac{1}{4}a^2\Lambda^2 = 0\, ,
\end{align} 
the solution of which is 
\begin{align}\label{critpts}
\Lambda^2 = \frac{2}{a^2}\left[ 
(n_0^2-\frac{a}{2}z) - \left( n_0^4 -a z n_0^2 - \frac{1}{4}a^2 x^2 \right)^{1/2} \right]\, .
\end{align}
The square root branch point defines a caustic surface, 
\begin{align}
 n_0^4 -a z n_0^2 - \frac{1}{4}a^2 x^2  =0\, ,
\end{align}
shown in figure \ref{fig:TwoRays}.
The four critical points in the complex $\Lambda$ plane are  real in the illuminated zone,  complex in the shadow zone and coalesce in pairs at the caustic.


\begin{figure}[!h]
	\center{
		\includegraphics[width= 200pt]{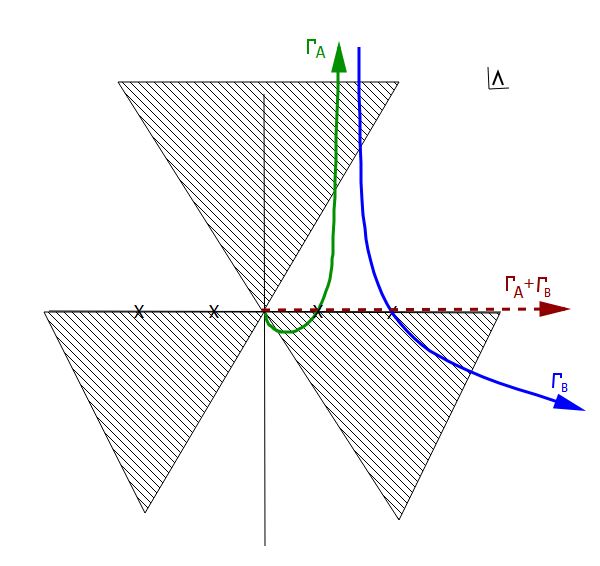}
		\caption{Steepest descent  contours or Lefschetz thimbles $\Gamma_A$ and $\Gamma_B$ in the complex $\Lambda$ plane, for $\vec x$ in the illuminated zone.  These contours pass through real critical points. The critical point associated with $\Gamma_A$ maps to the eigenray $A$  of figure \ref{fig:TwoRays} which has not touched the caustic, whereas the critical point along $\Gamma_B$ maps to the eigenray $B$ which intersects the caustic. The Greens function is a sum of integrals of $\Psi$ along these contours, equivalent to an integral along the positive real axis. The shaded wedges denote the angular domains at infinity within which integrals of $\Psi$ converge.}
		\label{fig:Contours1}
	}
\end{figure}

\begin{figure}[!h]
	\center{
		\includegraphics[width= 200pt]{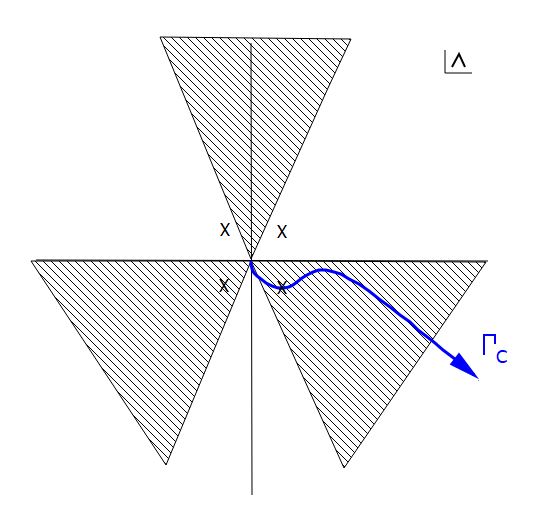}
		\caption{A steepest descent integration contour, or Lefschetz thimble, $\Gamma_C$ in the complex $\Lambda$ plane for $\vec x$ in the shadow zone.  The contour is equivalent to the positive real axis, and passes through a complex critical point, marked with  'X'. The shaded wedges denote the angular domains within which contours extending to infinity yield convergent integrals.}
		\label{fig:Contours2}
	}
\end{figure}



The Lefschetz thimbles are particular representatives of equivalence classes of convergent integration contours, where the equivalence classes are determined by the essential singularities of $\exp(ik_0\bar S)$.   Convergence requires that the poles of $\bar S$ are approached within certain angular domains in the complex $\Lambda$ plane.  For example, the contour can only approach the pole at infinity, given by the $-\frac{a^2}{12}\Lambda^3$ term in the einbein action \eqref{einbein_action}, within the angular domains
\begin{align}
arg(\Lambda) = [\frac{\pi}{3},\frac{2\pi}{3}],\,\,  [\pi,\frac{4\pi}{3}],\,\, [\frac{5\pi}{3},2\pi]\, ,
\end{align}
for positive real $a$.
Similiarly the simple pole at $\Lambda=0$, with positive residue $\frac{\vec x^2}{4}$, can only be approached within the domain 
$-\pi<arg(\Lambda)<0$.
In the illuminated zone, the positive real axis  $\Gamma$ is equivalent to a sum over two Lefschetz thimbles $\Gamma = \Gamma_A + \Gamma_B$, shown in figure \ref{fig:Contours1}.  The contour $\Gamma_A$ connects the pole at $\Lambda =0$ to the pole at infinity within the angular domain $arg(\Lambda)=[\frac{\pi}{3},\frac{2\pi}{3}] $. This component is responsible for the inhomogeneous term in the Green's function equation because the residue of the $\Lambda=0$ pole vanishes when $\vec x-\vec x'=0$.   There is a map between $\Gamma_A$ and the eigenray between $\vec x'=0$ and $\vec x$ which does not touch the caustic.  This can be verified by comparing the value of the einbein action at the critical point on $\Gamma_A$ to the ray theory travel time $\omega T$.  One could also reach the same conclusion by noting that the contour $\Gamma_A$ is similar to the Lefschetz thimble of of the constant $n^2$ case, shown in figure \ref{fig:ConstIndex}, surviving in the $a\rightarrow 0$ limit in which the caustic disappears.  The well known phase shift of $-\pi/2$ for the ray touching the caustic is none other than the difference in the angle of the two Lefschetz thimbles as they pass through their respective real critical points.

Neglecting the contribution from $\Gamma_B$, integration over $\Gamma_A$ yields a Green's function, albeit not one satisfying the radiation condition.
The contour $\Gamma_B$ connects the third order pole at infinity to itself, approaching within two distinct angular domains, $arg(\Lambda)=[\frac{\pi}{3},\frac{2\pi}{3}]$ and $arg(\Lambda)=[\frac{5\pi}{3},2\pi] $, and ceases to converge for $a=0$.  The corresponding eigenray is that shown in figure \ref{fig:TwoRays} which touches the caustic once.  Taken on its own, $\Gamma_B$ corresponds to a source free solution of the Helmholtz equation.  


Moving $\vec x$ from the illuminated zone to the shadow zone,  where there is no longer a real eigenray, the real critical points associated with  $\Gamma_A$ and $\Gamma_B$ coalesce at the caustic and then separate, moving in opposite directions away from the real axis. The contour $\Gamma$ is then equivalent to a single Lefschetz thimble $\Gamma_C$, rather than the sum of two, passing through one of the complex critical points as it connects the essential singularity at the origin to the domain $arg(\Lambda) = [5\pi/3,2\pi]$ at infinity.  This contour is shown in figure \ref{fig:Contours2}. Comparison of figures \ref{fig:Contours1} and \ref{fig:Contours2} shows how, as the pair of saddle points  pinch together and then diverge, the two contours $\Gamma_A + \Gamma_B$ are replaced by the single contour $\Gamma_C$.  There is, again, a phase shift, in this case $-\pi/4$, related to the orientation of the Lefschetz thimble as it passes through the critical point.  
Passing between the shadow and illuminated zones, the homology or equivalence class of the integration contour remains the same, while the representative Lefschetz thimbles changes discontinuously.

Sufficiently far from the caustic, the leading term in a large $k_0$ expansion is obtained by expanding the einbein action in a Taylor series about each of the critical points, truncating at quadratic order:
\begin{align}
{\bf \bar{\cal S}} \approx {\bf \bar{\cal S}}_{I} + \frac{1}{2} {\bf \bar{\cal S}}^{''}_{I}( \Lambda - \Lambda_I) ^2 
\end{align} 
such that \eqref{LamEff} becomes
\begin{align}
G\approx\sum_{I} \frac{1}{4\pi  \Lambda_I}  \sqrt{ \frac{\pi}{i k_0 S^{''}_I}}  e^{ik_0{\bf \bar{\cal S}}_I}\, ,
\end{align}
where the index $I$ labels the critical point associated with each of the Lefschetz thimbles, $\Gamma_I$, whose sum is equivalent to the real contour $[0,+\infty]$. 
Fourier transforming $G$ with respect to $\omega = k_0/c_0$ yields the time dependent field,
\begin{align}
\phi(t,\vec x) = \int d\omega e^{-i\omega t}G \approx \sum_I \int d\omega e^{-i\omega (t-\frac{S_I}{c_0})}c_I
\end{align} 
which is the Green's function of the wave equation,
\begin{align}
\left(-\frac{n(\vec x)^2}{c_0^2}  \partial_t^2 + \vec \nabla^2\right)\phi = \delta(t)\delta^D(\vec x)\, .
\end{align}
The arrival times due to the delta function source are 
\begin{align}
t_I = \frac{1}{c_0}{\rm Re}(S_I)\, ,
\end{align}
which is, by definition, $1/c_0$ times the real part of the einbein action anywhere along the Lefschetz thimble labeled by $I$.  
Note that complex critical points included in the sum also give arrivals,  albeit with some temporal smearing which is related to the imaginary part of the einbein action at the critical point $\bar S_I$. 
The existence of arrivals due to complex eigenrays has been emphasized in \cite{Chapman}.


\section{Quadratic $n^2$}\label{schan}

Another example for which the einbein action can be given exactly is a simple model of a sound channel in two spatial dimensions;
\begin{align}
n(X,Z)^2 = n_0^2  -\alpha Z^2\, .
\end{align}
The Fock-Schwinger-Feynman represenation of the retarded Green's function is 
\begin{align}\label{PropXchan}
G(\vec x',\vec x)= &\int_0^\infty d\Lambda \int {\cal D} \vec X(\tau)  \,e^{ik_0 {\cal L}[\vec X]} \\ 
&{\cal L} = \int_0^1 \left(  \frac{1}{4\Lambda}\left( \frac{d  X}{d\tau} \right)^2  +   \frac{1}{4 \Lambda}\left( \frac{d  Z}{d\tau} \right)^2   + \Lambda (n_0^2-\alpha Z^2) \right) d\tau\, . \label{Om}
\end{align}
Because $n^2$ is quadratic in $\vec X$, the path integral over $\vec X(\tau)$ can still be carried out exactly.  The Euler Lagrange equations $\frac{\delta {\cal L}}{\delta \vec X(\tau)}=0$ are,
\begin{align}\label{chanpath}
\frac{d^2 X}{d\tau^2} &=0 \nonumber \\
\frac{d^2 Z}{d\tau^2} &+4\Lambda^2 \alpha z =0
\end{align}
having the solution,
\begin{align}
X&= A+B\tau \\
Z&= C \cos(2\sqrt{\alpha}\Lambda\tau + \Theta)\, .
\end{align}
Inserting this solution into ${\cal L}$, for initial points $\vec x'$ and final points $\vec x$, yields the einbein action,
\begin{align}\label{chanac}
{\bf \bar{\cal S}}(\Lambda) = \frac{ (x-x')^2 }{4\Lambda} + \Lambda n_0^2 + \sqrt{\alpha} 
\frac{ (z'^2 + z^2)\cos{ ( 2\sqrt{\alpha}\Lambda) } - 2 z' z }{2\sin{ (2\sqrt{\alpha}\Lambda )}}\, .
\end{align}
This result can also be found in \cite{Palmer} after some small changes in notation, along with the full einbein wavefunction
\begin{align}\label{ChanPsi}
\Psi(\Lambda) = \frac{k_0}{4\pi i \Lambda} \left( {\rm sinc}(2\sqrt{\alpha}\Lambda)\right)^{-1/2} 
e^{ik_0\bar S}\, . 
\end{align}

The residue of the pole of $\bar S$  at $\Lambda=0$ is $\vec x^2/4$, which is the universal result for a delta function source, independent of the index of refraction.
The novel feature here, not seen in the previous examples, is the presence of additional poles of $\bar S$ at 
\begin{align}\label{lotsapoles}
\Lambda = \frac{\pi n}{2\sqrt{\alpha}},\,\,\,\, n\ne 0\, 
\end{align}
for integer non-zero $n$.
Varying $\alpha$ from zero to a finite value, these poles enter from $\Lambda=\infty$.  In fact, poles enter or disappear at $\Lambda=\infty$ in general under deformations of $n(\vec x)$.  More will be said about this point later.   
The residues of these poles are 
\begin{align}
{\rm res}(n) = \frac{(z-(-1)^n z')^2}{4}\, .
\end{align}
The resemblance to the residue of the  $\Lambda=0$ pole is not an accident, as the form of the residue is highly constrained by \eqref{Schrod}.  Integrating over a Lefschetz thimble with a single endpoint at one of the poles in \eqref{lotsapoles} would yield a solution of the Helmholtz equation with an extended delta function source of the form $J=\delta(z-(-1)^n z')$. 
However, for the sum over Lefschetz thimbles which is equivalent to the positive real $\Lambda$ axis, all such inhomogeneous terms cancel; every such pole is the endpoint of one Lefschetz thimble and the starting point of another.  We will therefore refer to these poles as ``ghost" poles and their locus of vanishing residue as ghost sources.


The ghost poles are values of $\Lambda$ for which the Euler-Lagrange equations derived from the Lagrangian ${\cal L}(\vec X,\dot{\vec X},\Lambda)$, in this case \eqref{Om}, have no solution for generic Dirichlet boundary conditions on $\vec X$. A solution exists only for boundary values at which the residue of the pole vanishes, i.e. the ghost sources.
In the present example, the Euler Lagrange equations \eqref{chanpath} have no solution for $2\sqrt{\alpha}\Lambda=\pi n$ unless $Z(1) = (-1)^n Z(0)$, corresponding to the ghost poles and ghost sources of the einbein action  \eqref{chanac}. 

\begin{figure}[!h]
\center{
\includegraphics[width= 400pt]{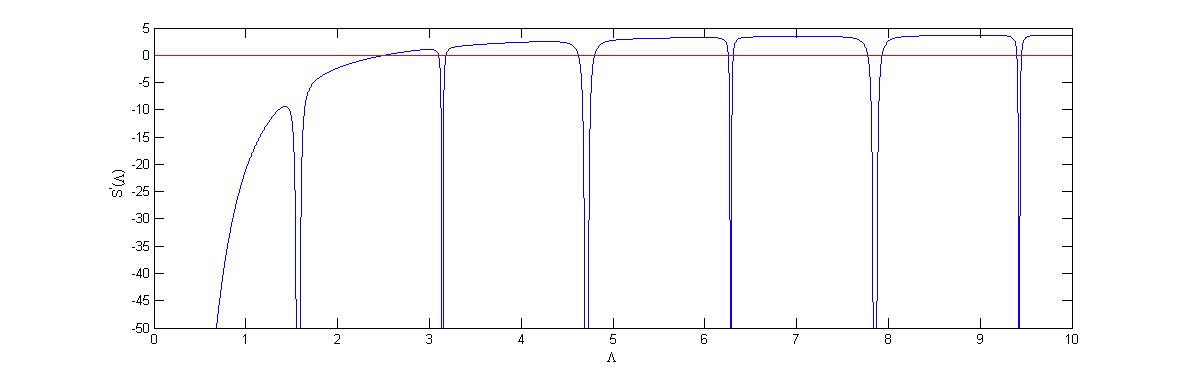}
\caption{$\frac{dS}{d\Lambda}$ shown as a function of real $\Lambda$.  The choice of other parameters leading to this curve is arbitrary, with only qualitative features of interest.  Real critical points are the intersections of this curve with the horizontal axis.  Upon varying parameters, such as $\vec x$, caustics occur when real critical points coalesce.}
\label{fig:ChannelCrit}
}
\end{figure}

The integral over the positive real $\Lambda$ axis is equivalent to an infinite sum of integrals over Lefschetz thimbles\footnote{This structure is almost captured in a contour integral representation shown in figure 4(b) of \cite{Holford1}, although the finite poles are circumvented.}  having endpoints at either the poles $\Lambda=\frac{\pi n}{2\sqrt{\alpha}}$ or at infinity in the upper half plane, each of which passes through a critical point at which 
\begin{align}
\frac{d\bar S}{d\Lambda} = -\frac{(x -x')^2}{\Lambda^2} + 
 n_0^2 - \alpha \frac{z'^2 +z^2 - 2z'z\cos(2\sqrt{\alpha}\Lambda)}{\sin^2(2\sqrt{\alpha}\Lambda)} = 0\, .
 \end{align}
 Figure \ref{fig:ChannelCrit} shows the structure of
 $\frac{d\bar S}{d\Lambda}$ for real $\Lambda$.
The infinite number of real critical points corresponds to eigenrays of ever steeper launch angle at the source $\vec x'$, having an arbitrary number of cycles. Increasing $x$ at fixed $z$ causes pairs of real critical points between the poles to merge at caustics, at which $\bar S''(\Lambda)=0$, and then become complex for increasing values of $\Lambda$. 


 A related example for which the einbein action is still exactly calculable is
\begin{align}
n^2(X,Z)= n_0^2 - \beta X- \alpha Z^2\, .
\end{align}
In this case,
\begin{align}
{\bf \bar{\cal S}}(\Lambda) = &\frac{ (x-x')^2 }{4\Lambda} + \Lambda (n_0^2+ \frac{\beta}{2}(x'+x) ) \nonumber \\
& + \sqrt{\alpha} 
\frac{ (z'^2 + z^2)\cos{ ( 2\sqrt{\alpha}\Lambda) } - 2 z' z }{2\sin{ (2\sqrt{\alpha}\Lambda )}}-\frac{1}{12}\beta^2\Lambda^3\, .
\end{align}
The finite poles are the same as in the previous example, however the pole at infinity is now third order rather than first, so the set of Lefschetz thimbles is larger.
There are still an infinite number of critical points, although the number of real critical points is now finite.

\section{Laurent series expansion in the einbein}\label{laurent}

It is useful to have methods to compute $\Psi(\Lambda)$ when $n^2(\vec X)$ is not quadratic. 
Perturbative computation of the the Fock--Schwinger--Feynman path integral using expansions of $n^2$ about the quadratic case is not the most powerful approach, as it is limited to small deformations.  A potentially more powerful method is proposed below, in which solutions of 
the Shr\"oedinger equation \eqref{Schrod} are obtained using a Laurent series expansion of the einbein action about finite $\Lambda$ poles.  
Each term in the Laurent series expansion is exactly calculable for a much larger class of $n^2$ than the quadratic subset. One can expand any given order in the Laurent series in a secondary expansion $1/k_0$.  Remarkably the $1/k_0$ expansion truncates. In fact, the first two terms in the Laurent series have no dependence on $k_0$.  The asymptotic nature of the $1/k_0$ expansion of solutions of the Helmholtz equation is apparent only after carrying out the integral $\int d\Lambda \Psi$. For reasons to be discussed,  Pad\'e approximants should be effective in extending the validity of the Laurent series far from any particular pole, capturing the existence of other poles which are intimately related to the existence of cusp caustics.




To solve the Schr\"oedinger equation \eqref{Schrod} by Laurent series expansion of $\bar S$ about the universal pole at $\Lambda=0$,  
one writes
\begin{align}
\Psi =  \left(\frac{k_0}{4\pi i \Lambda}\right)^{D/2} e^{ik_0\left( \gamma_{-1}\frac{1}{\Lambda} + \gamma_0+ \gamma_1 \Lambda + \cdots \right) }\, ,
\end{align}
such that \eqref{Schrod} becomes
\begin{align}
ik_0 \frac{\partial}{\partial\Lambda}\Psi(\Lambda)+ \left(\vec\nabla_x^2 + k_0^2n^2\right)\Psi(\Lambda) =
\left( 
\sum_{I=-2}^{\infty} \zeta_I \Lambda^I
\right)\Psi 
\end{align}
where $\zeta_l=0$ for all $l$. One can consistently take $\gamma_{0}=0$, which we do now for simplicity, although we will consider cases with non-vanishing $\gamma_0$ later.
At leading order, $\zeta_{-2}=0$ implies 
\begin{align}\label{nabsqd}
-(\vec \nabla_x \gamma_{-1})^2 + \gamma_{-1}=0\, . 
\end{align}
A solution of this equation is the residue of the universal pole, vanishing at the location of a source,
\begin{align}\label{lam2}
\gamma_{-1} = \frac{1}{4}(\vec x-\vec x')^2\, .
\end{align}
There are other solutions to \eqref{nabsqd}, involving different codimension sources, such as $\gamma_{-1} = \frac{1}{4}(z-z')^2$. However at the next order, $\zeta_{-1}=0$ implies
\begin{align}\label{srcdim}
\vec\nabla_x^2 \gamma_{-1}-\frac{D}{2} =0
\end{align}
which is satisfied by \eqref{lam2}.
Vanishing of $\zeta_0$ requires
\begin{align}
n(x)^2 - \gamma_1 - 2\vec{\nabla}_x\gamma_{-1}\cdot\vec{\nabla}_x\gamma_1 = 0
\end{align}
or using \eqref{lam2}
\begin{align}\label{gam1}
\gamma_1 = \left( 1+ (\vec x-\vec x')\cdot\vec{\nabla}_x\right)^{-1}n(x)^2\, .
\end{align}
Note that there is no $k_0$ dependence in any of the above expressions.  $k_0$ dependence arises at subsequent orders in the Laurent series expansion.  Requiring $\zeta_1=0$ to vanish gives
\begin{align}
-2 k_0^2 \vec{\nabla}_x\gamma_{-1}\cdot \vec{\nabla}_x\gamma_2 +ik_0\vec{\nabla}_x^2\gamma_1 - 2 k_0^2 \gamma_2 =0
\end{align}
or
\begin{align}\label{gam2}
\gamma_2 = \frac{i}{k_0}\left( 2+ (\vec x-\vec x')\cdot\vec{\nabla}_x\right)^{-1}\vec{\nabla}_x^2\gamma_1
\end{align}
with $\gamma_1$ given by \eqref{gam1}. This is the exact result for $\gamma_2$, valid in both the small and large $k_0$ limits. At next order, $\zeta_2=0$ implies
\begin{align}
-2k_0^2 \vec{\nabla}_x\gamma_{-1}\cdot\vec{\nabla}_x\gamma_3 - k_0^2(\vec{\nabla}_x\gamma_{1})^2 +ik_0\vec{\nabla}_x^2\gamma_2 - 3 k_0^2 \gamma_3 =0
\end{align}
or 
\begin{align}\label{gam3}
\gamma_3 = \left( 3+ (\vec x-\vec x')\cdot\vec{\nabla}_x\right)^{-1} \left(i\frac{1}{k_0}\vec{\nabla}_x^2\gamma_2 - (\vec{\nabla}_x\gamma_1)^2\right).
\end{align}
Using \eqref{gam1} and \eqref{gam2}, equation \eqref{gam3} yields an exact expression for $\gamma_3$, with one term independent of $k_0$ and the other scaling as $k_0^{-2}$. At a given order $m$ in the Laurent series in $\Lambda$, the $1/k_0$ expansion truncates at order $1/k_0^{(m-1)}$. 

The computation of the Laurent series simplifies considerably for an index of refraction which depends on a single variable. To illustrate a computation of the Laurent series explicitly, consider the index of refraction
\begin{align}
n^2= n_0^2 - AZ + BZ^3 \label{def1}
\end{align}
which for non-zero $B$ does not fall into the exactly soluble class described in the previous sections.  The solution of \eqref{gam1} is 
\begin{align}\label{solg}
\gamma_1 = \frac{1}{z-z'} \int_{z_0}^z dZ\, n^2(Z)
\end{align}
where one must choose $z_0=z'$ to obtain a non-singular result at $z=z'$.
Evaluating  \eqref{solg} for the case \eqref{def1} gives
\begin{align}
\gamma_1 = n_0^2 - \frac{A}{2}(z+z') + \frac{B}{4}(z+z')(z^2 + z'^2)\, .
\end{align}
Equation \eqref{gam2} yields,
\begin{align}
\gamma_2 &= \frac{i}{k_0}\frac{1}{(z-z')^2} \int_{z'}^z dZ (Z-z')\partial_{Z}^2 \gamma_1(Z) 
= \frac{iB}{2k_0}(z+z')\, .
\end{align}
From \eqref{gam3}, one has
\begin{align}
\gamma_3 &= \frac{1}{(z-z')^3} \int_{z'}^z dZ (Z-z')^2 (\frac{i}{k_0}\partial_{Z}^2 \gamma_2(Z) - (\partial_{Z}\gamma_1 )^2) \\      
= &-\frac{A^2}{12} +\frac{AB}{20}(3z^2 + 4zz'+3z'^2) - \frac{B^2}{112}(9z^4+20z^3z'+26z^2z'^2+20zz'^3+9z'^4)\, .
\end{align}
This procedure can be continued ad infinitum, and there is no reason to expect the Laurent series to truncate for non-zero $B$. 
Neglecting $1/k_0$ corrections and taking $z'=0$ for simplicity, 
\begin{align}
\bar S(\Lambda) = &\frac{1}{4\Lambda}\vec x^2 + (n_0^2 - \frac{A}{2}z+\frac{B}{4}z^3)\Lambda + \left(-\frac{A^2}{12} +\frac{AB}{20}(3z^2) - \frac{B^2}{112}(9z^4)\right)\Lambda^3 \nonumber\\ &+ \frac{B}{560}  \left(28 A^2z - 54 A B z^3+27 B^2 z^5\right)\Lambda^5 \nonumber \\
&+ \left( \frac{120120 A^3B - 848848A^2B^2z^2 + 1326780 A B^3z^4 - 601425B^4z^6}{16816800}\right)\Lambda^7 \cdots\, .
\end{align}
For $B=0$, one obtains the einbein action \eqref{einbein_action} for the linear $n^2$ case. Note that, to all orders, only $\gamma_{-1}$ has any dependence on coordinates other than $z$ when $n^2$ depends solely on $z$.

The potential existence of other poles at finite $\Lambda$ is not manifest in the Laurent series about the pole at $\Lambda = 0$.  Considering the example of section \ref{schan}, the exact wavefunction given by \eqref{chanac} and \eqref{ChanPsi} has poles at $\Lambda=\frac{\pi m}{2\sqrt{A}}$ for integer $m$.  One can write this wave function as 
\begin{align} \label{fullpole}
\Psi &= \frac{k_0}{4\pi i\Lambda} \prod_{m\ne 0}
\left( 
\frac{\pi m}{\pi m -2\sqrt{\alpha}\Lambda }
\right)^{1/2}e^{ik_0\bar S}\, ,  \\
\bar S &= \frac{ (x-x')^2 }{4\Lambda} + \Lambda n_0^2 + \frac{\sqrt{\alpha}}{2}(z^2 + z'^2)\sum_m \frac{1}{2\sqrt{\alpha}\Lambda - \pi m} - zz'
\frac{1}{2\Lambda}\prod_{m\ne 0}\frac{\pi m}{\pi m - 2\sqrt{\alpha}\Lambda}\label{chanac2}
\end{align}
where we have used the relations
\begin{align}
\sin(\theta) &= x\prod_{m=1}^\infty \left(1 - \frac{\theta^2}{\pi^2 m^2} \right) \\
\frac{\cos(\theta)}{\sin(\theta)} &= \sum_{m=-\infty}^{\infty}\frac{1}{\theta-\pi m}\, .
\end{align}
Comparing \eqref{chanac2} to the einbein action determined by Laurent series about $\Lambda=0$,
\begin{align} \Psi 
= \left(\frac{k_0}{4\pi i \Lambda}\right) e^{ik_0 (\gamma_{-1}\frac{1}{\Lambda} + \gamma_0+ \gamma_1 \Lambda + \cdots)} \, , \label{laurzero}
\end{align}
one concludes that the exponent of \eqref{laurzero} is the same as the Laurent series expansion of $ik_0\bar S'$ about $\Lambda=0$, where
\begin{align}
\bar S' \equiv \bar S + \frac{i}{k_0}  \sum_{m\ne 0}(1/2)\log\left( \frac{\pi m - 2\sqrt{\alpha}\Lambda}{\pi m} \right)\, .
\end{align}
Thus, for a solution based on a Laurent series,  the einbein action depends on the choice of the pole about which one expands, with 
results differing from a meromorphic einbein action such as \eqref{chanac2} by different logarithmic terms of order $1/k_0$.  

One can construct a Laurent series expansion about any of the poles, in the same manner as described above for the universal pole at $\Lambda=0$. For example, the  expansion of \eqref{fullpole} about the ghost pole $\lambda\equiv \Lambda - \frac{\pi}{\sqrt{\alpha}}=0$ has the form
\begin{align}
\Psi =  \left(\frac{k_0}{4\pi i \lambda}\right)^{1/2} e^{ik_0\left( \tilde\gamma_{-1}\frac{1}{\lambda} + \tilde\gamma_0+ \tilde\gamma_1 \lambda + \cdots \right) } \, .
\end{align}
The prefactor of the exponential behaves as $\lambda^{-1/2}$ rather than $\lambda^{-1}$ because the ghost source is co-dimension $D=1$, a curve in two spatial dimensions rather than a point.  In this case,
\begin{align}
\tilde\gamma_{-1} &= \frac{(z-z')^2}{4} \\
\tilde\gamma_0 &= \frac{ (x - x')^2}{ 4 \frac{\pi}{\sqrt A}}
\end{align}
with all higher terms  obtained iteratively from these using the Shr\"oedinger equation 
\begin{align}
ik_0 \frac{\partial}{\partial\lambda}\Psi(\lambda)+ \left(\vec\nabla^2 + k_0^2n^2\right)\Psi(\lambda) = 
\left(\sum_{I=-2}^\infty \zeta_{I} \lambda^I\right)\Psi  = 0 \, .
\end{align}
This yields results consistent with the exact solution, with an einbein action again differing from the meromorphic action \eqref{chanac2} by logarithmic terms of order $1/k_0$.

The Laurent expansion about any pole contains information about the other poles. The analysis above suggests a variant of Pad\'e approximants, where one starts from a Laurent series about the universal pole at $\Lambda=0$, which is then matched to an approximant of the form
\begin{align}\label{genpade}
\Psi \approx\left(\prod_m (\Lambda-\beta_m)^{-D_m/2}\right) \exp\left(ik_0\frac{A_N(\Lambda)}{B_M(\Lambda)}\right)\, ,
\end{align} 
where $A_N(\Lambda)$ and $B_M(\Lambda)$ are polynomials in $\Lambda$ of degree $N$ and $M$ respectively, $\beta_m$ are the zeros of $B_M$, and $D_m$ are the codimensions of the sources or ghost sources.  
The  large $\Lambda$ behavior of the exponent is $\bar S\sim\Lambda^{N-M}$, such that the pole at infinity is of degree $m_\infty = N-M$, and there are $m_\infty$ distinct angular wedges in which the $\Lambda$ integration can approach infinity.  There are $n_P = M$ simple poles at finite $\Lambda$ which also serve as enpoints of integration.  The number of critical points  of $\bar S$, or eigenrays, is 
\begin{align}
n_C= N+M-1 = m_\infty + 2n_P-1\, .
\end{align}
This relation is an example of the 
Riemann-Hurwitz formula, regarding $\bar S(\Lambda)$ as a map of the Reimann sphere onto itself.  For a degree ${\cal N}$ covering map of a Riemann surface of Euler characteristic $\chi$  onto a Riemann surface of Euler characteristic $\chi'$, the formula reads, 
\begin{align}\label{RH}
\chi ={\cal N}\chi' - \sum_i (m_i-1)
\end{align} 
where the integer $m_i$ is the ramification index at the i'th branch point of the inverse map $\bar S\rightarrow \Lambda$.  
In addition to the ramification index $m_\infty$ due to the behavior at infinity,  there is a ramification index $m_{i'}\ge 2$ due to each critical point at which $\bar S \sim (\Lambda - \Lambda_i)^{m_{i'}}$.  At caustics $m_{i'} \ge 3$. The degree $\cal N$ of a meromorphic map $\bar S(\Lambda)$ is the sum of the order of each pole, including that at infinity.  Thus for the map of the Reimann sphere to itself, with $\chi=\chi'=2$, \eqref{RH} becomes 
\begin{align}\label{critcount}
\sum_{i'} &(m_{i'}-1) = 2{\cal N}-2 - (m_{\infty}-1) \, , \nonumber \\
 &{\cal N} = m_\infty + n_P 
\end{align}
Away from caustics, the left hand side of \eqref{critcount} is simply the number of critical points.

There is evidence, to be discussed later, that the counting of cusp caustics is related to $n_P$.
Thus it may be possible to choose an optimal approximant, i.e. choose $N$ and $M$ in \eqref{genpade}, based on prior knowledge of critical points and cusp caustics.
The underlying assumption on which the power of the Pad\'e approximation depends is that singularities of $\mathbb S = -i\ln\Psi(\Lambda)$ are only poles or logarithms, where the finite poles are simple.  There are no other singularities in the exactly soluble cases with quadratic $n^2$.   
It is not clear what physical interpretation other singularities might have, although we cannot exclude them, whereas the simple poles are intimately related to ghost sources, cusp caustics and monodromies as will be shown later.  Note that the Schr\"oedinger equation requires logarithmic singularities of $\mathbb S$ coincide with the poles of  $\bar S$.


Given some perturbation of $n^2(\vec x)$,  one can readily write integral expressions for the corresponding effect on terms in the Laurent series.
The reader may wonder if one can compute motion of the poles under perturbations, but this is not the case; only the relative positions of poles is determined by solving the Schroedinger equation, which is invariant under $\Lambda \rightarrow \Lambda+ \Delta$.
The fact that the poles may move relative to each other, or that new poles may appear under perturbation, is not manifest from the Laurent expansion about a particular pole, but could be visible in the generalized Pad\'e approximant.  

Consider a perturbation which is turned on continuously, $n^2 \rightarrow n^2 + \delta  { \Omega}(\vec x)$, with $\delta$ evolved from $0$ to $1$. In doing so, new poles can enter from infinity, but may not appear spontaneously at finite $\Lambda$ having a vanishing residue at finite $\vec x$.  If a new pole did appear spontaneously,  it would have the local form 
\begin{align}\label{spont}
\bar S = \frac{f(\delta,\vec x)}{\Lambda -\beta}+\cdots
\end{align}
with $f(0,\vec x)=0$.  However the form of residue is highly constrained by equations such as \eqref{nabsqd} and \eqref{srcdim}, admitting no small solutions and having no explicit dependence on $n(\vec x)$.
There is a way that new poles can appear by  splitting poles at finite $\Lambda$, but this does not occur under deformations of the index of refraction.  Instead pole splitting occurs via the smearing of delta function sources in a manner described in section \ref{cusp}.  Remarkably, this splitting is accompanied by the formation of a cusp caustic.

 \section{Splitting poles by source smearing: \\ 
constant $n^2$ with a cusp caustic}\label{cusp}

The poles of the einbein action depend on both the index of refraction and the source.  
Here we describe deformations of the source, from a point supported delta function to an extended curve, resulting in the splitting of poles and the formation of a cusp caustic. 

Consider a constant sound speed, two spatial dimensions and a source, 
\begin{align}\label{src1}
J(\vec x) = \delta(z) \exp\left(-ik_0\frac{x^2}{4\mu} \right)\, .
\end{align}
Using the results of section \ref{const}, the solution of the Helmholtz equation,
\begin{align}
\left(\vec\nabla^2 + k_0^2 n(\bf x)^2\right)\phi = J(\vec x)
\end{align}
is given by the einbein integral
\begin{align}\label{CuspInt}
\phi(\vec x)=\int\, &d\vec x' G(\vec x',\vec x)J(\vec x) = \int_0^\infty d\Lambda \Psi(\Lambda) \nonumber \\
 \Psi &=
\frac{1}{4\pi  \Lambda}  \int dx' e^{-ik_0\frac{x'^2}{4\mu}} e^{ik_0 \left(\frac{1}{4\Lambda}(x- x')^2 + \frac{1}{4\Lambda}z^2 +\Lambda n_0^2\right)} \nonumber \\
&=   \sqrt{\frac{ i\mu}{4\pi k_0\Lambda(\Lambda-  \mu)}}
e^{ik_0 \bar S}
\end{align}
where 
\begin{align}\label{spole}
\bar S \equiv   \frac{1}{4(\Lambda- \mu)} x^2 +\frac{1}{4\Lambda}z^2  +\Lambda n_0^2\, .
\end{align}
Thus the deformation parameterized by $\mu$ gives rise to a splitting of the pole at $\Lambda=0$.  This pole, with residue $(x^2 + z^2)/4$, splits into a pole at $\Lambda=0$ with residue $z^2/4$ and a pole at $\Lambda=\mu$ with residue $x^2/4$.  The latter is necessarily a ghost pole, since the locus on which its residue vanishes does not correspond to the domain of support of the source \eqref{src1}.   As will be seen later, there are two contributing Lefschetz thimbles bounded by the ghost pole, such that there is no additional inhomogeneous term in the Helmholtz equation at $x=0$. 

The manner in which new poles can appear is highly constrained by \eqref{Schrod}, for reasons discussed in section \ref{laurent}.  In particular, if the residue $\gamma_{-1}$ vanishes over a codimension $d$ surface $\Sigma$, it satisfies the equations,
\begin{align}\label{poleconstraint}
-(\vec \nabla \gamma_{-1})^2 + \gamma_{-1}=0 \nonumber \\
\vec\nabla^2 \gamma_{-1}-\frac{d}{2} =0\, .
\end{align}
The solution of \eqref{poleconstraint} is
\begin{align}\label{polform}
\gamma_{-1} = \frac{1}{4} \vec\xi^2\, ,
\end{align} 
where $\vec\xi$ are coordinates transverse to $\Sigma$, with distance to the source given by $\vec\xi^2$.  
Under deformation of the source, equation \eqref{poleconstraint} allows the pole associated with $\Sigma$  to split into other poles having orthogonal surfaces $\Sigma_I$ of vanishing residue, subject to a constraint on their codimensions $\sum_I d_I = d$.  

Although we give no examples here, ghosts sources could lie on curved or complicated domains.  Writing 
\begin{align}\label{foliation}
d\vec x^2 = g_{ij}(\xi,\vec\sigma)d\sigma^id\sigma^j + d\xi^2\, ,
\end{align}
where the  ghost source has support on a curved surface at $\vec\xi=0$ parameterized by $\vec \sigma$,   the residue of the associated pole is still given by \eqref{polform}.  However higher order terms in the Laurent expansion of the Shr\"oedinger equation for $\Psi$ could impose 
constraints on the geometry of ghost sources.   If \eqref{foliation} is a singular foliation of space, expressions for higher order terms in the Laurent expansion, similar to 
\eqref{gam1} and \eqref{solg}, may be ill defined in a manner analogous to ambiguities due to intersecting characteristic curves.

\begin{figure}[!h]
\center{
\includegraphics[width= 200pt]{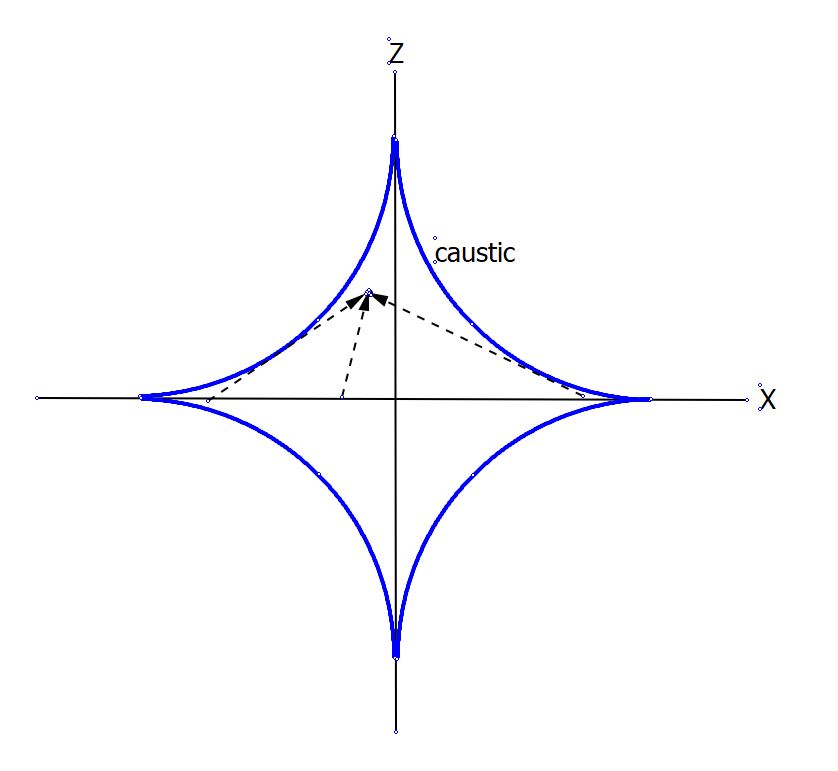}\caption{Three real ray paths for $\vec x$ within the cusp caustic.}
\label{fig:cusprays}
}
\end{figure}

\begin{figure}[!h]
\center{
\includegraphics[width= 200pt]{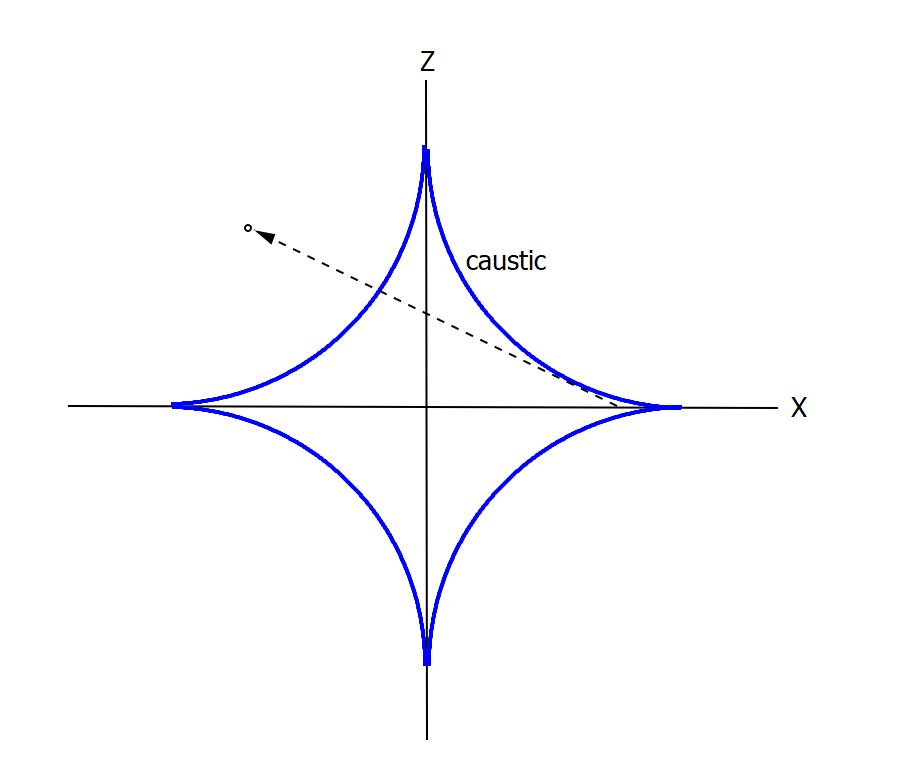}
\caption{Single real ray path for $\vec x$ outside the cusp caustic.}
\label{fig:cusprayout}
}
\end{figure}

\begin{figure}[!h]
\center{
\includegraphics[width= 200pt]{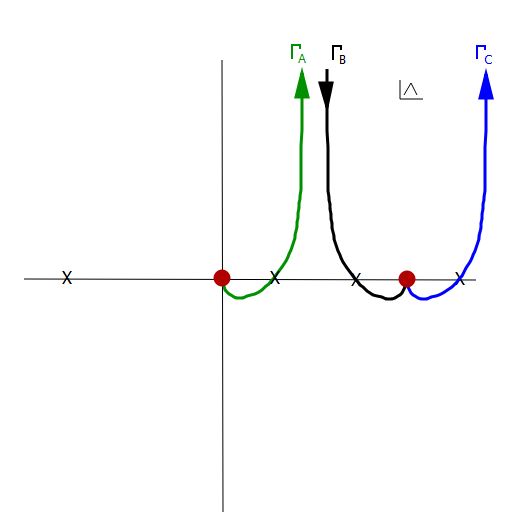}
\caption{Poles, critical points and Lefschetz thimbles for $\vec x$ within the cusp caustic.  Poles are indicated by circles and critical points by X.  The pole on the right is a ghost pole at $\Lambda=\mu$ with residue $\frac{1}{4}x^2$, bounding two Lefschetz thimbles. Each of the Lefschetz thimbles marked $\Gamma_A,\Gamma_B,\Gamma_C$ correspond to the real rays shown in figure \ref{fig:cusprays}. The Lefschetz thimble $\Gamma_A$ maps to the ray which does not intersect the caustic. The bounding pole of $\Gamma_A$ at $\Lambda=0$ gives rise to the source term \eqref{src1}.}
\label{fig:incusp}
}
\end{figure}

\begin{figure}[!h]
\center{
\includegraphics[width= 200pt]{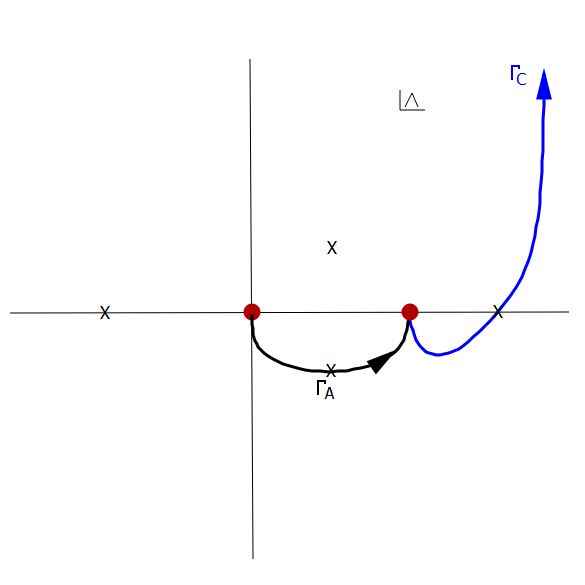}
\caption{Poles, critical points and Lefschetz thimbles contributing to the solution for $\vec x$ outside the cusp caustic.  Poles are indicated by circles and critical points by X.  The pole on the right is a ghost pole, bounding two  Lefschetz thimbles. The Lefschetz thimble marked $\Gamma_C$ corresponds to the real ray shown in figure \ref{fig:cusprayout}. The Lefschetz thimble $\Gamma_A$ maps to a complex ray.}
\label{fig:outcusp}
}
\end{figure}

The Riemann-Hurwitz formula implies that the appearance of a new pole is necessarily accompanied by new critical points.     
For the present example \eqref{spole},  a single pole at $\Lambda=0$  becomes two poles at finite $\Lambda$ for non-zero deformation $\mu$, with the pole at infinity unchanged.  Instead of two critical points, there are now four,  shown in figures \ref{fig:incusp} and \ref{fig:outcusp}.  The deformation parameterized by $\mu$ also gives rise to a caustic with a cusp.  Upon crossing this caustic, two of the critical points transition from real to complex.
The  caustic surface is given by  
\begin{align}
x^{2/3} + z^{2/3} = \left(2n_0\mu\right)^{2/3}\, ,
\end{align}
with cusp singularities at $(x,z)= (0, \pm 2n_0\mu)$.
The unit vector describing the launch angle of rays at the $z=0$ source is 
\begin{align}
[\hat n_x,\hat n_z] = \left[ -\frac{X}{2\mu},\,\, \sqrt{1-\left( \frac{X}{2\mu}\right)^2} \right] 
\end{align}
such that three rays reach any point inside the cusp, $x^{2/3} + z^{2/3} < (2\mu n_0)^{2/3}$, as shown in figure \ref{fig:cusprays}, while only one real ray reaches points outside as shown in figure \ref{fig:cusprayout}.  Inside the caustic surface, there are three Lefschetz thimbles contributing to the solution $\phi$, each of which passes through a real critical point as shown in figures \ref{fig:incusp}.  Outside the cusp,  there are only two contributing Lefschetz thimbles,  passing through one complex critical point and one real critical point as shown in figure \ref{fig:outcusp}.

The pole of the einbein action \eqref{spole} at $\Lambda=0$ corresponds to the source,  whereas that at $\Lambda = \mu$ is a ghost pole.  Since one Lefschetz thimble ends and another begins at the ghost pole, as can be seen in figures \ref{fig:incusp} and \ref{fig:outcusp},   there is no inhomgeneous term in the Helmholtz equation with support at the points where the ghost pole residue vanishes. 
Only one Lefschetz thimble is bounded at $\Lambda=0$, such that the surface term in the integral of the Shroedinger equation \eqref{Schrod} is  
\begin{align}\label{surfterm2}
\left(\vec\nabla^2 + k_0^2n^2\right)\phi = -ik_0  \int_\Gamma d\Lambda \frac{\partial}{\partial\Lambda}\Psi = -ik_0\Psi_{i0^-} ,
\end{align}
where $\Psi_{i0^-}$ denotes the limit of $\Psi$ as it approaches $\Lambda=0$ from below on the imaginary axis.
Given \eqref{CuspInt} and \eqref{spole}, one obtains
\begin{align}
\Psi_{i0^-} = \delta(z) \exp\left(-ik_0\frac{x^2}{4\mu} \right)=J(\vec x)\, .
\end{align}

Under rather general conditions, nearby pairs of poles give rise to cusp caustics, even if the pair was not generated by source deformation.  The einbein action in the neighborhood of a sufficiently close pair of poles can be written as 
\begin{align}\label{nearpoles}
\bar S = \frac{r_1^2}{4(\Lambda - p - \Delta/2)} + \frac{r_2^2}{4(\Lambda - p+\Delta/2)} + Q(\Lambda)
\end{align}
where $r_{1,2}$ are distances to the sources or ghost sources, and $Q$ is well approximated by a linear function of $\Lambda$ between the poles,
\begin{align}\label{nearpoles2}
Q \approx a + b(\Lambda-p)\, .
\end{align}
The caustic consists of points $\vec x$ at which both first and second derivatives of the einbein action vanish simultaneously. Determining this surface is essentially the same calculation for the action given by \eqref{nearpoles} and \eqref{nearpoles2} as for \eqref{spole}, yielding 
\begin{align}
r_1^{2/3}+r_2^{2/3} = (2 \sqrt{b}\Delta)^{2/3}\, .
\end{align}
So long as $r_1$ and $r_2$ may be independently varied by changing $\vec x$, the caustic has the same shape as in figure \ref{fig:cusprayout} when displayed in $r_1,r_2$ space. 
Mapping to position space will yield some deformation of this shape, but the cusp singularity persists.

 \section{Uniform asymptotic approximations to the einbein integral in the neighbohood of a smooth caustic}\label{unif}

The uniform asymptotic approximation \cite{Pearcey,Chester,Kravtsov1,Kravtsov2,Ludwig,Berry,Berry2,Berry3,Holford2}  is a well known method  to extend the domain of validity of ray theory beyond the illuminated zone far from a caustic.  The results are typically accurate at large $k_0$ in the immediate neighborhood of a caustic, including some small distance into the shadow zone.  Below we derive uniform asymptotics for a smooth caustic in the language of the einbein. In essence, the einbein action is approximated by a third order Taylor expansion about a suitable point $\Lambda_c$, such that the integral is can be expressed in terms of an Airy function and its derivatives. More general non-smooth caustics will be discussed in section \ref{class}.

Consider the Laurent series expansion of the einbein action in the neighborhood of some pole $P$, truncated at order $(\Lambda-P)^3$;
\begin{align}\label{3trunc}
{\bf \bar{\cal S}}(\Lambda) &\approx \gamma_{-1}\frac{1}{\tilde\Lambda} + \gamma_0+ \gamma_1 \tilde\Lambda + \gamma_2\tilde\Lambda^2 +\gamma_{3} \tilde\Lambda^3\, , \nonumber\\ &\tilde \Lambda \equiv\Lambda-P\, .
\end{align}
This can in turn be Taylor expanded about a point $\tilde\Lambda_c(\vec x,\vec x')$ chosen such that the second order term vanishes,
\begin{align}\label{unifax}
\bar S &\approx \Gamma_0 + \Gamma_1\lambda +  \Gamma_3 \lambda^3 + \cdots\nonumber \, ,\\ 
&\lambda \equiv  (\tilde\Lambda - \tilde\Lambda_c)\, .
\end{align}
In this approximation, $\bar S(\Lambda)$ has a pair of critical points at which $d\bar S/d\Lambda=0$.  Moving $\vec x$ towards a smooth caustic,  the two critical points coalesce at $\lambda=0$, at which $d^2 \bar S/d\Lambda^2=0$.
The critical points are a pair of $\lambda$ of opposite signs which are real in the illuminated zone and imaginary in the shadow zone.
Note that truncation at third order is insufficient to describe a higher order caustic, at which more than two critical points coalesce. 
In fact, approaching a cusp point, three critical points converge on a ghost pole, rather than a point at which $\bar S(\Lambda)$ is regular.  This phenomenon will be discussed in the subsequent section.

\begin{figure}[!h]
	\center{
		\includegraphics[width= 200pt]{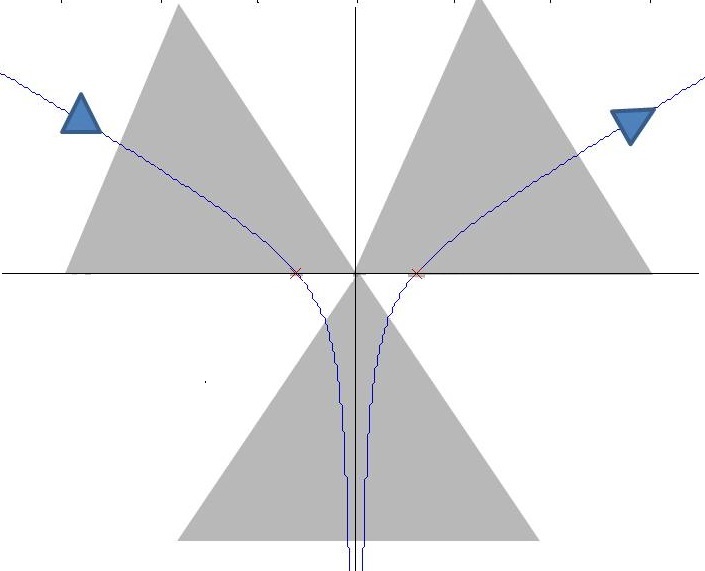}
		\caption{Lefschetz thimbles in the complex $\lambda$ plane, for $\vec x$ in the illuminated zone, derived from the uniform asymptotic approximation of the einbein action given in \eqref{unifax}.  These pass through  real critical points marked with 'X'.  The Greens function is given by the sum of an integral along these contours. The shaded wedges denote the angular domains within which contours extending to infinity yield convergent integrals.}
		\label{fig:Contours3}
	}
\end{figure}

\begin{figure}[!h]
	\center{
		\includegraphics[width= 200pt]{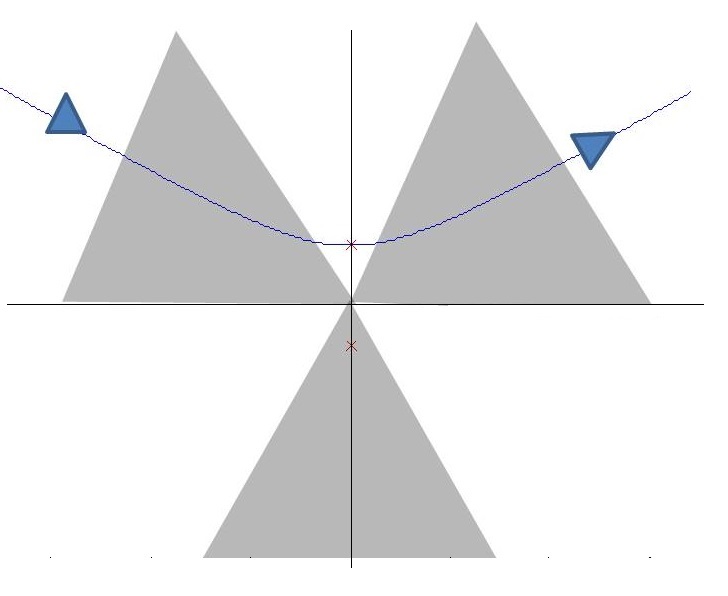}
		\caption{A Lefschetz thimble in the complex $\lambda$ plane, for $\vec x$ in the shadow zone, derived from the uniform asymptotic approximation of the einbein action given in \eqref{unifax}. This contour passes through a complex critical point, marked with  'X'.  The Green's function is given by the integral along this contour. The shaded wedges denote the angular domains within which contours extending to infinity yield convergent integrals.}
		\label{fig:Contours4}
	}
\end{figure}

If the pole $P$ is associated with a codimension $d$ source, or ghost source, a Lefschetz thimble $\Gamma_I$ with endpoint on $P$ yields a contribution $\phi_I$ to the field $\phi$ of the form 
\begin{align}\label{uniform}
\phi_I &\approx \frac{i}{k_0}\int_{\Gamma_I} d\lambda \left(\frac{1}{4\pi (\lambda+\Lambda_c)}\right)^{d/2} e^{ i\left( \Gamma_0 + \Gamma_1\lambda + \Gamma_3 \lambda^3 \right) } \nonumber\\
&\approx  \frac{i}{k_0}\frac{1}{(4\pi)^{d/2}}\int_\Gamma d\lambda \left(
\tilde\Lambda_c^{-d/2} - \lambda \frac{d}{2}\tilde\Lambda_c^{-1-d/2}  + {\cal O}(\lambda^2\cdots) 
\right) e^{ i\left( \Gamma_0 + \Gamma_1\lambda + \Gamma_3 \lambda^3 \right) }\, .
\end{align}
Note that a Lefschetz thimble in the complex $\lambda$ plane, derived from \eqref{unifax},  is an approximate construct.  It is not equivalent to the Lefschetz thimble determined from the full einbein action using the definition $\lambda\equiv \Lambda-P-\tilde\Lambda_c$. 
Having Taylor expanded about $\tilde\Lambda = \tilde\Lambda_c$,  the essential singularity at $\tilde\Lambda=0$ is not apparent.  Furthermore, the terms in the Taylor expansion of $\bar S$ which come from the pole, expanding $1/\tilde\Lambda$ about $\tilde\Lambda=\tilde\Lambda_c$, alter the coefficient of the cubic term.   In fact the sign of the cubic term changes generically.  
The domains at infinity over which the approximate integral \eqref{uniform} converges are completely different from that of the exact einbein integral.  However, the behavior of the integrand near the critical points is the essentially the same.
In crossing from the illuminated zone to the shadow zone,  the Lefschetz thimbles of the approximate integral collapse from a pair passing through two real critical points, to a single contour passing through an imaginary critical point, as shown in figures \ref{fig:Contours3} and \ref{fig:Contours4}.  This process mimics the evolution of the Lefschetz thimbles of the exact solution.  

The integral \eqref{uniform} is given in terms of the Airy function and its derivatives, since
\begin{align}
\int_\Gamma dt\, t^n \exp(ut + \frac{1}{3}t^3)= \frac{d^n}{du^n} Ai(u)\, .
\end{align}
For the two spatial dimension example of section \ref{sec3}, with $n^2=n_0^2 - az$ and a source $J=\delta^2(x,z)$,
one obtains the uniform asymptotics,
\begin{align}
G&\approx \frac{1}{4\pi  \Lambda_c}e^{i\left(-2\gamma_{3}\Lambda_c^3 + \gamma_{1}\Lambda_c\right) } 2\pi\left[{\rm Ai}\left(\frac{6\gamma_3\Lambda_c^2+\gamma_1}{(-6\gamma_3)^{1/3}}\right)  - \frac{1}{\Lambda_c(-6\gamma_3)^{1/3}} {\rm Ai}^{'} \left(\frac{6\gamma_3\Lambda_c^2+\gamma_1}{(-6\gamma_3)^{1/3}}\right)  +\cdots\right] \nonumber \\
&\Lambda_c=\left(\frac{-\gamma_{-1}}{3\gamma_3}\right)^{1/4}
\end{align}
with
\begin{align}
\gamma_{-1} &= \frac{1}{4}(x^2 + z^2) \nonumber \\
\gamma_1 &= n_0^2 - \frac{a}{2}z \nonumber \\
\gamma_3 &= -\frac{1}{12}a^2 \, .
\end{align}



\section{Einbein action and Thom--Arnold classifcation}
\label{class}

In the neighborhood of a general caustic, solutions of the Helmholtz equation have a uniform asymptotic approximation with an integral representation of the form
\begin{align}
\phi(\vec x) \approx \int d\lambda_1 \cdots d\lambda_n\,  e^{ik_0 P \left[ {\bf \lambda_1 \cdots \lambda_n},{\bf \zeta_1(\vec x)\cdots\zeta_K}(\vec x) \right]}\, ,
\end{align} 
where $\zeta_J$ parameterize the space transverse to the caustic and $P({\bf \lambda},{\bf\zeta})$ is a polynomial in $\lambda_i$ of the form
\begin{align}
P({\mathbf \lambda},{\bf\zeta})=P_0({\bf \lambda}) + \sum_{J=1}^K \zeta_J Q_J(\lambda)\, .
\end{align}
The caustic surface is the locus of $\zeta$ where critical points satisfying $\frac{\partial P}{\partial \lambda}=0$ coalesce via degeneration of the matrix $\frac{\partial^2 P}{\partial \lambda_i \partial\lambda_j}$.
 The polynomial  $P$ is  the generating function in the classification of catastrophes due to Thom \cite{Thom} and Arnold \cite{Arnold}. Examples known as the elementary catastrophes are shown in table \ref{AD_catast}.  The smooth caustic corresponds to the $A_2$ catastrophe, for which the relation between the einbein action and the generating polynomial was shown in section \ref{unif}.  

\begin{figure}[!h]
	\center{
		\includegraphics[width= 420pt]{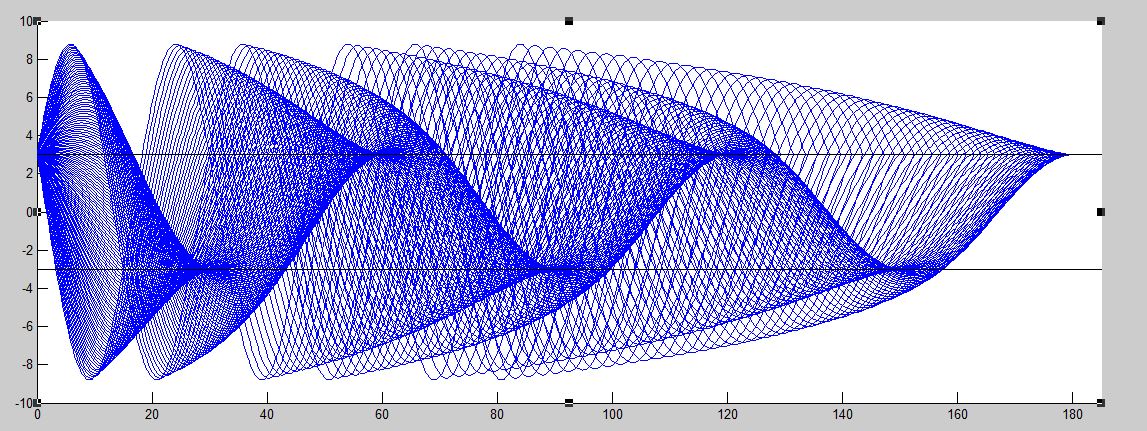}
		\caption{Fan of rays for  the index of refraction $n^2=n_0^2-\alpha Z^2$ considered in section \ref{schan}, with  $\alpha=0.01, n_0=1$ and source at $x'=0,z'=3$.  The rays are generated from a set of evenly spaced launch angles between $\pm pi/3$.  The caustics are manifest as surfaces where the rays become dense.  The cusp singularities of the caustic curves occur exactly at the location of ghost sources, $z=\pm z'$, shown as horizontal lines overlaying the fan of rays.}
		\label{fig:RayFan}
	}
\end{figure}

The map between the einbein action and $A_N$ catastrophes with $N>2$ is more subtle than that for $A_2$.  
For the caustic described in section \ref{cusp}, the cusp or $A_3$ catastrophe lies at a spatial point at which the smooth components of a caustic intersect a ghost source. We conjecture that this is generally true; cusp caustics always occur within the domain of support of ghost sources. This can also be seen for the example of quadratic $n^2$ described in section \ref{schan}.  The cusps can be seen clearly in the fans of rays shown in figure \ref{fig:RayFan}, and all lie on the surfaces $z=z'$ corresponding to ghost sources.   

In the neighborhood of an $A_N$ caustic
one can define a variable
\begin{align}\label{mapP}
\lambda \equiv \nu(\vec x)\left(\Lambda-\tilde{\Lambda}(\vec x)\right)
\end{align}
such that, for a certain choice of functions $\nu(\vec x)$ and $\tilde\Lambda(\vec x)$, the einbein action is approximated by a polynomial of the form
\begin{align}\label{poly}
\bar S&\approx P(\lambda,\vec \zeta)= \frac{\lambda^{N+1}}{N+1} + \zeta_{N-1} \frac{\lambda^{N-1}}{N-1} + \zeta_{N-2} \frac{\lambda^{N-2}}{N-2} +\cdots+\zeta_1\lambda\, .
\end{align}
However the map $\Lambda\rightarrow\lambda$ is singular when $\vec x$ is the location of a ghost source.  Approaching the cusp $\zeta_1(\vec x)=\zeta_2(\vec x)=0$ of the caustic described in section \ref{cusp}, three critical points of $P$ in the complex $\lambda$ plane coalesce at $\lambda=0$; the corresponding behavior in the complex $\Lambda$ plane involves three critical points of the einbein action converging on the ghost pole at $\Lambda=\mu$. The residue of the pole vanishes just as the three critical points collide which occurs, by definition, within the domain of support of a ghost source.  This process is illustrated in figure \ref{fig:cuspcritmapsingular}.

\begin{figure}[!h]
	\center{
		\includegraphics[width= 150pt]{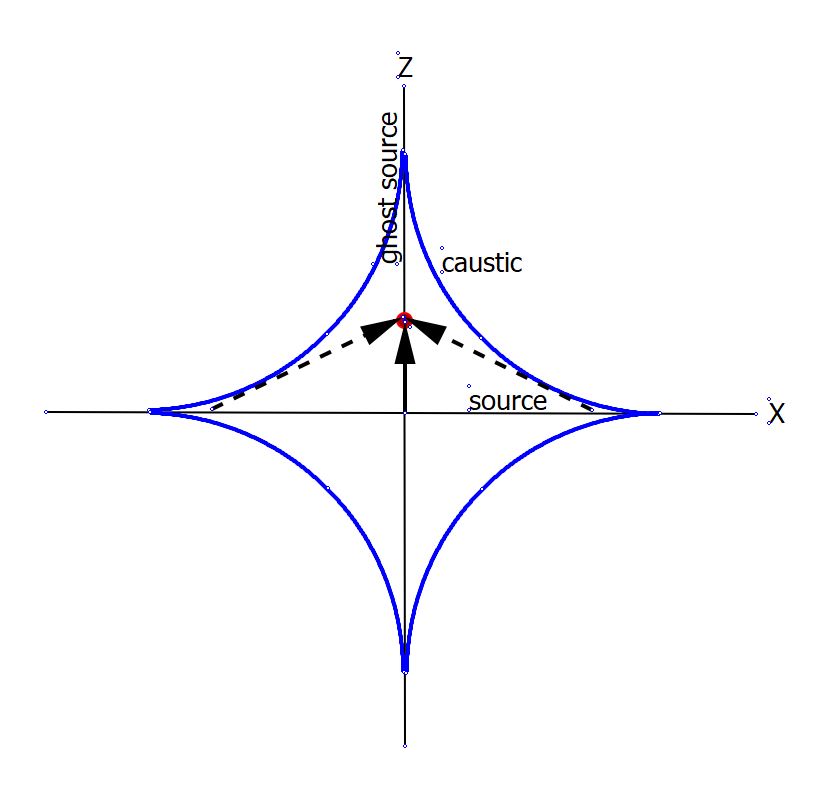}
		\caption{Within the cusp and along the ghost source at $x=0$, there are two distinct arrivals.  The first arrival is due to the ray emanating from the origin. The second is due to the pair of rays which touch the caustic and reach points along $x=0$ simultaneously.}
		\label{fig:two_arrivals}
	}
\end{figure}

\begin{figure}[!h]
	\center{
		\includegraphics[width= 150pt]{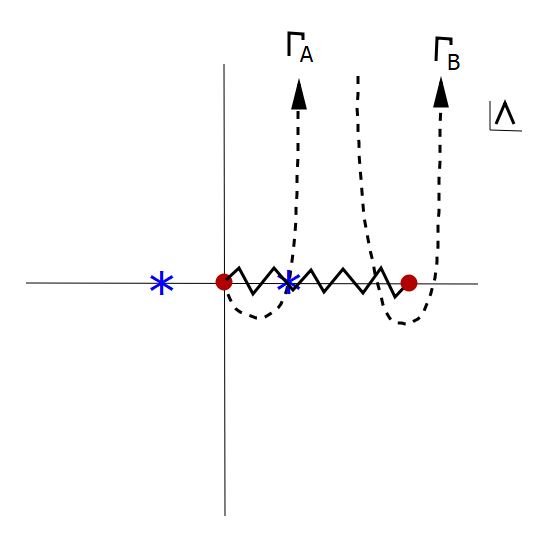}
		\caption{The field at $x=0$, along the ghost source, is obtained by summing the integral of the wave function $\Psi(\Lambda)$ over the contours $\Gamma_{A}$ and $\Gamma_B$.  The critical points are indicated by *. Although the einbein action $\bar S$ only has a pole at $\Lambda=0$ for points along $x=0$, there are branch points of $\Psi$ at $\Lambda=0$ and $\Lambda=\mu$.  The contour $\Gamma_B$ does not pass through a critical point of $\bar S$, but instead winds around the branch point at $\Lambda=\mu$, connecting the essential singularity at infinity on one Riemann sheet to the essential singularity at infinity on the other Riemann sheet.}
		\label{fig:BranchPt}
	}
\end{figure}

Since the generating function $P(\lambda)$ has no poles, the map \eqref{mapP} is singular  at a ghost source.  In a special case considered in \cite{Holford1}, the mas constructed using methods described in \cite{Bleistein}.
Even far from a caustic,  varying $\vec x$ across a ghost source exhibits interesting behavior of critical points of the einbein action in the complex $\Lambda$ plane, whereas nothing interesting occurs in the complex $\lambda$ plane of the generating function.
In the $\Lambda$ plane, this process is accompanied by 
a `sterile' collision of critical points, for which there is no corresponding collision in the complex $\lambda$ plane.  Real pairs of critical points appear to cross through each other at the location of a ghost pole, remaining real since there is no caustic.  Matlab code generating an animation of this process for $\vec x$ crossing both caustics and ghost sources, in the context of the einbein action of section \ref{cusp}, is given in appendix \ref{appendixB}, along with figures illustrating snapshots at certain spatial points.  Construction of the map $\Lambda\rightarrow\lambda$ at general $A_{N>2}$ caustics is an interesting task which will not be attempted here.

 For $\vec x$ on the ghost source,
The einbein action \eqref{spole} becomes,
\begin{align}\label{reduced}
\bar S = \frac{1}{4\Lambda}z^2 + \Lambda n_0^2\, .
\end{align}
which has two fewer critical points than $\bar S$ at $\vec x \ne 0$, away from the ghost source.  There is no corresponding loss of critical points in the generating function $P(\lambda)$, reflecting the singularity of the map \eqref{mapP}.  
Approaching the ghost source,  
\begin{align}
\frac{d\lambda}{d\Lambda} &= \nu(\vec x)\rightarrow \infty  \nonumber \\
&\tilde\Lambda(\vec x) \rightarrow \Lambda_{c}^-
\end{align}
where $\Lambda_c^- = \frac{z}{2n_0}$ is the surviving critical point of \eqref{reduced} for which there is a corresponding critical point of the generating function. 
The loss of two critical points at the ghost source may seem puzzling, as it naively suggests a single arrival in the temporal picture (the Fourier transform with respect to $k_0$). Yet there are clearly two distinct arrival times at the ghost source $x=0$ within the cusp $z<2n_0\mu$, as illustrated in figure \ref{fig:two_arrivals}.  The resolution lies in the fact that $\Psi$ still has a branch point singularity at $\Lambda=\mu$, even when the residue of the ghost pole of $\bar S$ has vanished. As written previously in  \eqref{CuspInt} and \eqref{spole},
\begin{align}\label{persist}
 \Psi =
  &\sqrt{\frac{ i\mu}{4\pi k_0\Lambda(\Lambda-  \mu)}}
e^{ik_0 \bar S}
 \nonumber \\
&\bar S =   \frac{1}{4(\Lambda- \mu)} x^2 +\frac{1}{4\Lambda}z^2  +\Lambda n_0^2\, .
\end{align}
The wave function $\Psi$ has branch points at $\Lambda=0$ and $\Lambda=\mu$, where the latter persists even when $x=0$ and there is no longer an essential singularity at $\Lambda=\mu$.
It follows that the integral representation $\phi=\int d\Lambda \Psi$ involves two contours rather than one. One of these begins at the essential singularity at $\Lambda=0$ and passes through the critical point $\Lambda=\frac{z}{2n_0}$ of \eqref{reduced} before continuing to the essential singularity at infinity.  The other contour begins at essential singularity at infinity, wraps around the branch point at $\Lambda=\mu$ and then continues back to infinity on the other Riemann sheet, as shown in figure \ref{fig:BranchPt}.  This contour corresponds to the second arrival which is not apparent solely from consideration of the einbein action \eqref{reduced}.


\begin{figure}[!h]
	\center{
		\includegraphics[width= 150pt]{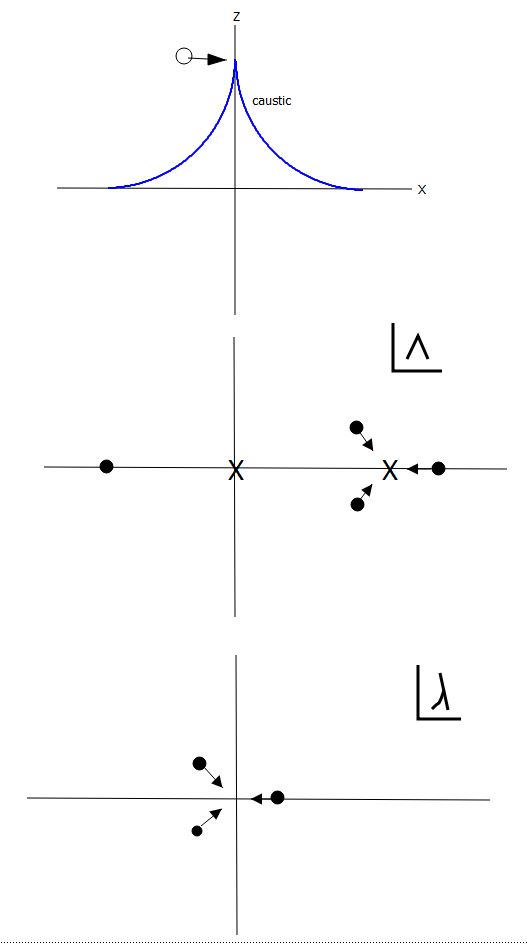}
		\caption{Approaching the singular point of a cusp caustic for the example of section {cusp}, together with the motion of the critical points of the einbein action $\bar S(\Lambda)$ in the complex $\Lambda$ plane and the critical points of the generating function $P(\lambda)$ in the complex $\lambda$ plane. In the complex $\Lambda$ plane, the critical points converge upon the ghost pole. No such pole is apparent in the $\lambda$ plane.  The critical point on the far left in the complex $\Lambda$ plane is essentially a spectator,  with no associated contributing Lefschetz thimble, as well as  no corresponding critical point of the generating function $P(\lambda)$. }
		\label{fig:cuspcritmapsingular}
	}
\end{figure}

\begin{table}
	\begin{center}
		\begin{tabular}{|l|l|l|}
			\hline
			$A_2$ & Smooth or Fold Caustic & $\frac{1}{3}\lambda^3 + \zeta_1\lambda$ \\ 
			\hline
			$A_3$ & Cusp Caustic & $\frac{1}{4}\lambda^4 + \frac{1}{2}\zeta_2 \lambda^2 + \zeta_1\lambda$ \\
			\hline
			$A_4$ & Swallowtail Caustic & $\frac{1}{5}\lambda^5 + \frac{1}{3}\zeta_3\lambda^3 + \frac{1}{2}\zeta_2\lambda^2 + \zeta_1\lambda$ \\
			\hline
			$A_5$ & Butterfly Caustic & $\pm\frac{1}{6}\lambda^6 + \frac{1}{4}\zeta_4\lambda^4 + \frac{1}{3}\zeta_3\lambda^3 + \frac{1}{2}\zeta_2\lambda^2 + \zeta_1\lambda$ \\
			\hline
			$D^+_4$ & Hyperbolic Umbilic & $\lambda_1^3 + \lambda_2^3 - \zeta_3\lambda_1\lambda_2-\zeta_2\lambda_2-\zeta_1\lambda_1$ \\ 
			\hline
			$D_4^-$ & Elliptic Umbilic & $\lambda_1^3 - 3\lambda_1\lambda_2^2-\zeta_3(\lambda_1^2+\lambda_2^2)-\zeta_2\lambda_2-\zeta_1\lambda_1$ \\
			\hline
			$D_5$ & Parabolic Umbilic & $\lambda_1^4+\lambda_1\lambda_2^2+\zeta_4\lambda_2^2+\zeta_3\lambda_1^2 + \zeta_2\lambda_2 + \zeta_1\lambda_1$ \\
			\hline	
		\end{tabular}\caption{Some caustics and their generating functions.}\label{AD_catast}
	\end{center}
\end{table}

There are other catastrophes besides the $A_N$ type, whose generating function is polynomial in two variables $\lambda_{1,2}$. 
Some examples of this type are the $D_N$ and $E_6,E_7,E_8$   catastrophes.  There are also more exotic catastrophes having moduli, or variable numeric parameters in their generating functions, which have more than two $\lambda_i$.  One could say that there must be more than one einbein for all caustics besides the $A_N$ type. It is not clear how such cases can ever occur for an analytic index of refraction, for which the Fock-Schwinger-Feynman representation involves one degree of freedom $\Lambda$.  On the other hand scattering problems have solutions involving pairs of Greens functions coupled by a scattering kernel. Perhaps scattering giving rise to the  $D_N$ and $E_{6,7,8}$ catastrophes have representations involving two coupled einbein, with uniform asymptotics mapping  onto the known generating functions.


 \section{Monodromies}\label{monodromies}

Solutions of the Helmholtz equation have branch point singularities at certain points in the space of parameters defining the index of refraction, as well as at source locations in coordinate space. Analytic continuation in closed loops around the branch points mixes solutions in a manner characterized by the monodromy group. This group can be determined simply from the singularities of the einbein wave function $\Psi(\Lambda)$, without knowledge of the full solution $\phi = \int d\Lambda \Psi$.  These singularities determine the convergent integration contours in the complex $\Lambda$ plane.  Thus
as parameters of the index of refraction, or coordinates $x_i$, are varied around a closed complex loop, the integration contours in the complex $\Lambda$ plane vary to maintain convergence, mixing amongst themselves non-trivially. 
 
    
In general, the einbein action has the large $\Lambda$ behavior 
\begin{align}
\bar S \sim \mu \Lambda^P
\end{align}
such that convergence of $\int d\Lambda\Psi$ requires approaching infinity within angular wedges for which ${\rm Im}(\mu \Lambda^P)>0$. These wedges rotate cyclically into each other as one analytically continues $\mu$ in a closed loop about $\mu=0$. 
The finite poles of the einbein action all have the form,
\begin{align}\label{finwedge}
\bar S \sim \frac{\xi^2}{4(\Lambda-\beta)}\, ,
\end{align}
and can only be approached within wedges for which the imaginary part of \eqref{finwedge} is positive.  For real $\xi$, the wedge is 
$\pi <arg(\Lambda-\beta)<2\pi$.  These wedges are fixed under variations of the parameters defining the index of refraction, but rotate under analytic continuation of the coordinates $\xi$. Varying the parameters defining the index of refraction around closed loops,  integration contours change in a way constrained by the convergent direction of approach to the bounding poles of $\bar S$. In the subsequent discussion, example of monodromy groups are given for a few examples.
One can also think of the monodromy group as relating different eigenrays,  since contours in different homology classes  map to eigenrays. 

\begin{figure}[!h]
	\center{
		\includegraphics[width= 200pt]{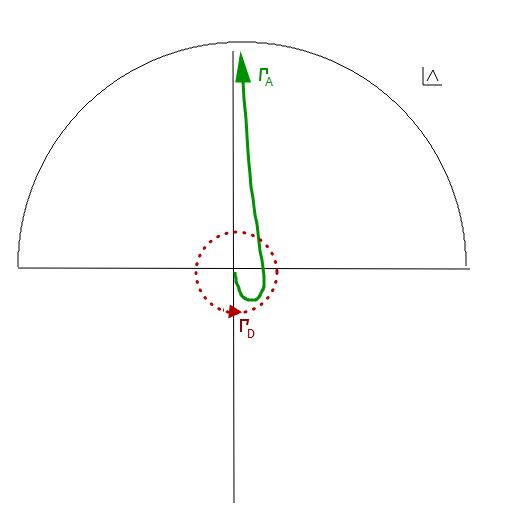}
		\caption{Contours in the complex $\Lambda$ plane corresponding to a basis set of linearly independent solutions which are closed under monodromy with respect to $\nu\equiv n_0^2$, in two spatial dimensions. These particular contours are arbitrarily chosen representatives of homology classes, rather than Lefschetz thimbles.}
		\label{fig:Monodromies2}
	}
\end{figure}

\begin{figure}[!h]
	\center{
		\includegraphics[width= 200pt]{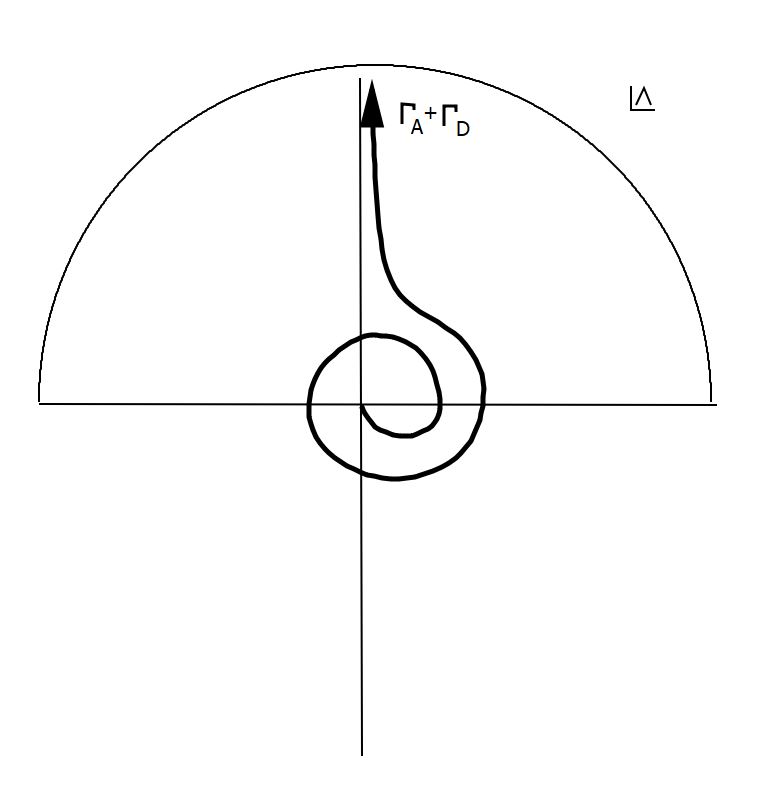}
		\caption{Contour in the complex $\Lambda$ plane obtained from the action of the monodromy on $\Gamma_A$, under which $\Gamma_A\rightarrow\Gamma_A+\Gamma_D$. Due to the essential singularity at $\Lambda=0$, which must be approached from the lower half plane in order for the integral to converge,  one can not use Cauchy's integral theorem to unwind the loops around the origin and recover $\Gamma_A$. This contour is an arbitrarily chosen representative of a homology class, rather than a Lefschetz thimble.}
		\label{fig:Monodromies4}
	}
\end{figure}

As a very simple example,   consider the case $n(\vec x)^2 = n_0^2 \equiv \nu$ in two spatial dimensions, described in section \ref{const}.  For a source at $\vec x'=0$,
\begin{align}
\Psi = \frac{k_0}{4\pi i \Lambda}e^{
	ik_0\left( \frac{1}{4\Lambda}|\vec x|^2  + \nu\Lambda\right)
}\, .
\end{align}
At large $\Lambda$, $\bar S \sim \nu\Lambda$, so that the convergence wedge at $\Lambda\rightarrow\infty$ is $0<arg(\Lambda)<\pi$ for real $\nu$. We wish to determine the monodromy as one continuously varies $\nu \rightarrow \exp(2\pi i)\nu$, under which the convergence wedge at infinity undergoes a $2\pi$ rotation.  A basis set of contours $\Gamma_{A,D}$ closed under the monodromy is shown in figure \ref{fig:Monodromies2}. The contour $\Gamma_D$, closed around the essential singularity at $\Lambda=0$, corresponds to a standing wave solution: a sum of ingoing and outgoing waves centered at $\vec x=0$. Defining $\tilde\Gamma_A$ to be the reflection of $\Gamma_A$ about the imaginary axis, $\Gamma_D = \Gamma_A-\tilde\Gamma_{A}$. The essential singularity at $\Lambda=0$ can only be approached from ${\rm Im}(\Lambda) <0$.  Thus, as one rotates the phase of $\nu$, and with it the convergence wedge at large $\Lambda$, a contour ending on $\Lambda=0$ is forced to wind around $\Lambda=0$ as shown in figure \ref{fig:Monodromies4}. This reflects the fact that the two dimensional Greens function, the Hankel function $H^{(1)}_0(\sqrt{\nu}|\vec x|)$, has an infinite number of Riemann sheets in $\nu$ with a branch point at $\nu=0$. 
The monodromy is
\begin{align}
\begin{pmatrix}
\Gamma_A \\ \Gamma_D 
\end{pmatrix}
\rightarrow
\begin{pmatrix}
1 & 1 \cr
0 & 1
\end{pmatrix}
\begin{pmatrix}
\Gamma_A \\ \Gamma_D 
\end{pmatrix}\, .
\end{align}

\begin{figure}[!h]
	\center{
		\includegraphics[width= 200pt]{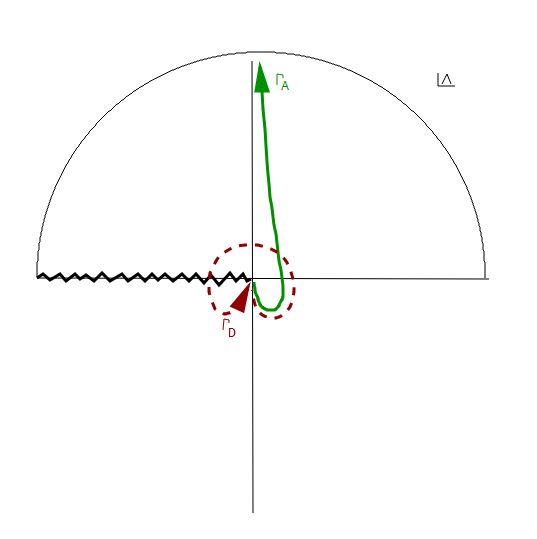}
		\caption{Contours in the complex $\Lambda$ plane corresponding to a basis set of linearly independent solutions which are closed under monodromy with respect to $\nu\equiv n_0^2$, in three spatial dimensions.  In three dimensions $\Lambda=0$ is both an essential singularity and branch point of $\Psi(\Lambda)$. These particular contours are arbitrarily chosen representatives of  homology classes, rather than Lefschetz thimbles.}
	\label{fig:Monodromies3}
	}
\end{figure}

\begin{figure}[!h]
	\center{
		\includegraphics[width= 200pt]{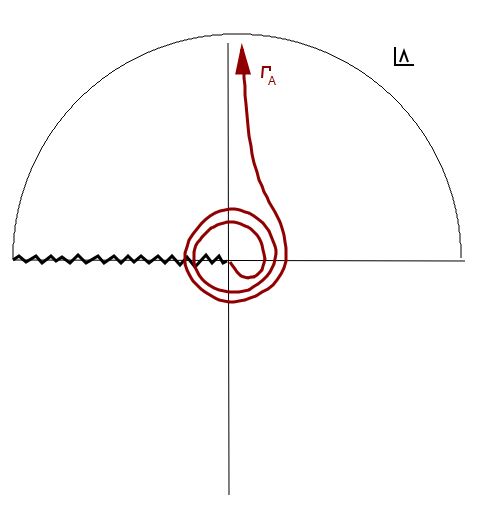}
		\caption{Contour obtained after acting twice with the monodromy starting with $\Gamma_A$.  Due to the sign change between the two Riemann sheets, acting twice is equivalent to the identity: $M^2:\Gamma_A\rightarrow\Gamma_A$. This contour is an arbitrarily chosen representative of a homology class, rather than a Lefschetz thimble.}
		\label{fig:Monodromies5}
	}
\end{figure}

In three dimensions, the Greens function 
$\exp( ik_0 \sqrt{\nu}|\vec x|)/(4\pi|\vec x|)$ has only two Riemann sheets in $\nu$. In this case,
\begin{align}
\Psi = \left(\frac{k_0}{4\pi i \Lambda}\right)^{3/2}e^{
	ik_0\left( \frac{1}{4\Lambda}|\vec x|^2  + \nu\Lambda\right)
}
\end{align}
such that the essential singularity at $\Lambda=0$ coincides with a square root branch point which was not present in the two dimensional case. A basis set of contours closed under the monodromy is shown in figure \ref{fig:Monodromies3}. The contour $\Gamma_D$, again corresponding to a standing wave, begins at the essential singularity at $\Lambda=0$, crosses to the second Reiman sheet in $\Lambda$ and then ends at $\Lambda=0$.  Starting with the contour $\Gamma_A$ and continuously varying $\nu\rightarrow \exp(4\pi i)\nu$ yields the contour in figure \ref{fig:Monodromies5}, which is equivalent to $\Gamma_A$ because of cancellations due to the sign difference between the Riemann sheets in $\Lambda$.  The monodromy  is
\begin{align}
\begin{pmatrix}
\Gamma_A \\ \Gamma_D 
\end{pmatrix}
\rightarrow
\begin{pmatrix}
-1 & 1 \cr
0 & 1
\end{pmatrix}
\begin{pmatrix}
\Gamma_A \\ \Gamma_D 
\end{pmatrix}\, .
\end{align}   
The square of the monodromy matrix is the identity, reflecting the existence of only two Riemann sheets.

\begin{figure}[!h]
	\center{
		\includegraphics[width= 250pt]{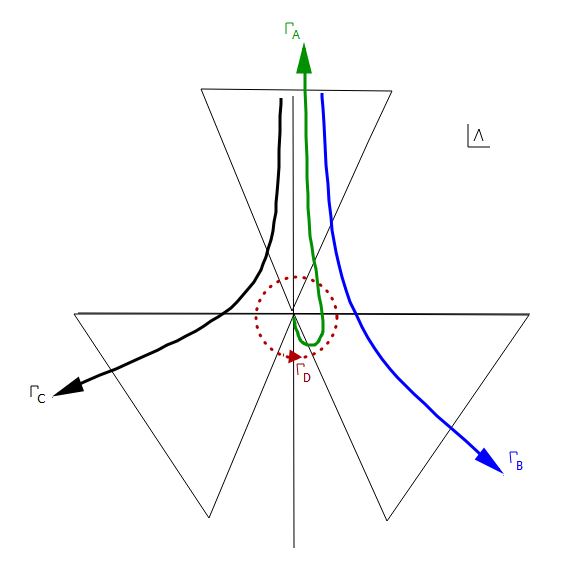}
		\caption{Contours in the complex $\Lambda$ plane corresponding to a basis set of linearly independent solutions which are closed under monodromy with respect to closed loops about $A=0$, in two spatial dimensions. These contours are  arbitrarily chosen representatives of homology classes, rather than Lefschetz thimbles.}
		\label{fig:Monodromies}
	}
\end{figure}

 More complicated monodromies occur upon considering non-constant $n^2(\vec x)$.  
Consider the two dimensional example of section \ref{sec3}, with $n^2(\vec x)=n_0^2 - a z$ and source at $\vec x'=0$ such that 
\begin{align}
\Psi = \frac{k_0}{4\pi i \Lambda}e^{ik_0\left(\frac{1}{4\Lambda}\left(x^2 + z^2\right) + \Lambda\left( \nu - \frac{ a}{2}z\right) - \frac{1}{12}a^2\Lambda^3\right)}\, .
\end{align}
There is no longer a branch point at $\nu=0$, since convergence of $\int d\Lambda\Psi$ at infinity is insensitive to the argument of $\nu$. Instead there is  a branch point at $a=0$.  A non-trivial monodromy arises upon varying the argument of $a$ by $2\pi$. 
A basis set of contours $\Gamma_{A,B,C,D}$ which is closed under the monodromy is shown in figure \ref{fig:Monodromies}. The contour $\Gamma_A+\Gamma_B$ corresponds to the radiation condition Green's function.  The contour $D$, closed around the essential singularity of $\Psi$ at $\Lambda=0$, yields a standing wave solution.  
This singularity can only be approached from ${\rm Im}(\Lambda) <0$.  Thus, as one rotates the argument of $a$, and with it the convergence wedges at large $\Lambda$, a contour ending at both $\Lambda=0$ and $\Lambda=\infty$  may be forced to wind around $\Lambda=0$.  Rotating the convergence wedges at $\Lambda\rightarrow\infty$ along with the argument of $a$, while keeping the convergence wedge at $\Lambda=0$ fixed, leads to the monodromy 
\begin{align}\label{monofour}
\begin{pmatrix}
\Gamma_A \\ \Gamma_B \\ \Gamma_C \\ \Gamma_D 
\end{pmatrix}
\rightarrow M \begin{pmatrix}
\Gamma_A \\ \Gamma_B \\ \Gamma_C \\ \Gamma_D 
\end{pmatrix},\,\,\,\,\,\,\,\, M=
\begin{pmatrix}
1 & 1 & 0 & 0 \cr
0 & -1 & 1 & -1 \cr
0 & -1 & 0 & 0 \cr
0 & 0 & 0 & 1
\end{pmatrix}
\, .
\end{align}   
In the illuminated zone, $\Gamma_A$ is associated with the ray which does not touch the caustic, while  $\Gamma_B$ is associated with the ray which touches the caustic once.  Thus the monodromy \eqref{monofour} implies that these rays are related to each other by analytic continuation of $a$. 
Varying the argument of $x^2+z^2$ by $2\pi$, such that the convergence wedge at $\Lambda=0$ rotates by $2\pi$, yields the monodromy,
\begin{align}
\begin{pmatrix}
\Gamma_A \\ \Gamma_B \\ \Gamma_C \\ \Gamma_D 
\end{pmatrix}
\rightarrow
\begin{pmatrix}
1 & 0 & 0 & 1 \cr
0 & 1 & 0 & 0 \cr
0 & 0 & 1 & 0 \cr
0 & 0 & 0 & 1
\end{pmatrix}
\begin{pmatrix}
\Gamma_A \\ \Gamma_B \\ \Gamma_C \\ \Gamma_D 
\end{pmatrix}=
 M^{-3} \begin{pmatrix}
\Gamma_A \\ \Gamma_B \\ \Gamma_C \\ \Gamma_D 
\end{pmatrix}
\, .
\end{align}

For ghost poles, a $2\pi$ rotation of the argument of $\xi^2$ in \eqref{finwedge} has no effect. 
The two contours ending and beginning at a ghost pole obtain  winding contributions upon varying the argument of $\xi^2$. However, since the orientations of these winding components are opposite, they  cancel. Indeed, solutions of the Helmholtz equation do not have singularities at the location of ghost sources\footnote{There are singularities of caustic curves appearing within the domain of support of ghost sources, but these are not singularities of the solution.}.

\begin{figure}[!h]
	\center{
		\includegraphics[width= 300pt]{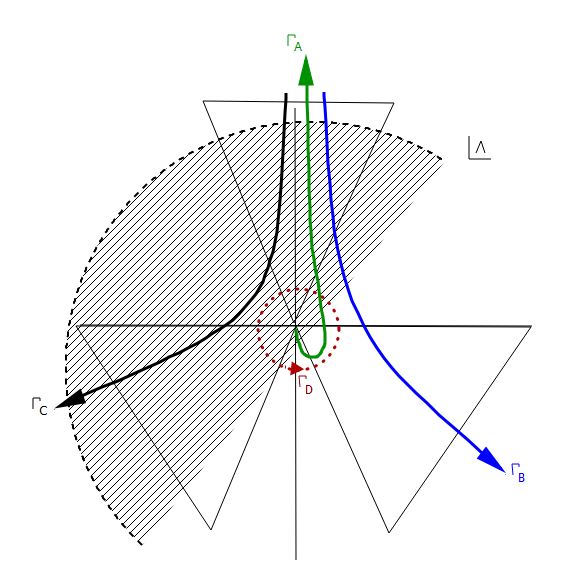}
		\caption{A basis set of contours closed under monodromy as in figure \ref{fig:Monodromies}, except that $n_0^2$ is complex so that the convergence domain at infinity for $a=0$ lies within the semicircular region shown, corresponding to a rotation of the upper half plane.  The integral over the contour $\Gamma_C$ now vanishes in the $a\rightarrow 0$ limit, since it is contractible within this domain.} 
		\label{fig:Monodromies6}
	}
\end{figure}

For non-zero $a$, there is no branch point at $\nu=0$.  The appearance of multiple Riemann sheets in $\nu$ as $a\rightarrow 0$  has an explanation given for a mathematically similar problem in \cite{GG1}. As $a\rightarrow 0$, the solution set collapses; some contours with finite $\int d\Lambda \Psi$  vanish or diverge in the limit, depending on the argument of $\nu$.  Although convergence is insensitive to $arg(\nu)$ at finite $a$, existence of the $a\rightarrow 0$ limit is dependent on $\arg(\nu)$.  
While varying $\arg(\nu)$ at finite $a$, one can at some point add the contribution due to a contour $\Gamma'$ for which the $a\rightarrow 0$ limit vanishes, 
\begin{align}\label{addGam}
G=\frac{i}{k_0}&\int_\Gamma d\Lambda\Psi \rightarrow \frac{i}{k_0}\int_{\Gamma+\Gamma'} d\Lambda\Psi\, , \nonumber \\
&\lim_{a\rightarrow 0} \int_{\Gamma'} d\Lambda \Psi=0\, . 
\end{align}  
An example of a contour for which the integral vanishes as $a\rightarrow 0$ is shown in figure \ref{fig:Monodromies6}.  Although \eqref{addGam} would constitute an abrupt change in boundary conditions for non-zero $a$, it is irrelevant as $a\rightarrow 0$.  Continuing to vary $\arg(\nu)$ however, the addition of $\tilde\Gamma$ becomes important as one crosses a stokes line.  In fact $\Gamma'$ can be chosen such that the the integral remains finite and is analytic in $\nu$ as $a\rightarrow 0$.    Continuing this process while varying $arg(\nu)$ from $0$ to $2\pi$ yields the non-trivial monodromy described above. 

Another class of monodromies arises upon  variations of the source function $J(\vec x)$.  Moving the endpoints of the integration contour away from essential singularities is equivalent to considering smooth sources, rather than delta functions, given by
\begin{align}
J(\vec x) = -ik_0(\Psi(\Lambda^+) - \Psi(\Lambda_-))\, .
\end{align}
Thus varying endpoints in non-trivial closed loops in the complex $\Lambda$ plane maps to non-trivial closed loop variations of the source.

\section{Concluding remarks}

The intent of this article has been a detailed analysis of the analytic structure of the Fock--Schwinger--Feynman proper-time solution of the  Helmholtz equation.  This solution is a specific gauge fixing of a more general path integral formulation in which the `proper time' $\Lambda$  is a gauge fixed einbein.  Integrating the spatial path degrees of freedom first gives rise to the einbein action formulation considered here, whereas integrating the einbein first gives a formulation whose large $k_0$ expansion is related to ray theory.
Ordinarily expressed as an integral of a wave function $\Psi(\Lambda)$ over the  positive real $\Lambda$ axis, the Fock--Schwinger--Feynman  solution can be formulated in terms of 
a sum over steepest descent paths $\Gamma_i$, or  Lefschetz thimbles, in the complex $\Lambda$ plane, bounded by essential singularites of the the integrand $\Psi(\Lambda)=\exp(i{\mathbb S}(\Lambda))$.   The function $\Psi$ contains an enormous amount of information about the solution, which is manifest even before the integration over $\Lambda$ is carried out. The essential singularities of $\Psi$ corresponding to poles of ${\mathbb S}$ are intimately related to a number of phenomena including sources, eigenray genesis under perturbation, cusp caustics, and monodromies. 

We have often referred to the leading term in the large $k_0$ expansion of $\mathbb S$ as the einbein action $\bar S$,  whose critical points map to eigenrays.  In the excactly soluble case, $\bar S$ is the meromorphic part of $\mathbb S$, differing from $\mathbb S$ by logarithmic terms.  The real part of the einbein action
is a constant along each Lefschetz thimble $\Gamma_i$, equal to the arrival time for a temporal delta function source. 
The critical points of the einbein action are  minima  of ${\rm Im}(\bar S)$ along $\Gamma_i$, at which $\bar S$ is a solution of the eikonal equation corresponding to real or complex eigenrays. Complex critical points are of particular importance in shadow zones,  and even in illuminated zones where they give rise to arrivals neglected by standard approaches based on real rays. 
The number and topology of Lefschetz thimbles changes discontinuously upon crossing a caustic, at which critical points coalesce.   
Although eigenray based methods are considered legitimate at large $k_0$, the sum over Lefschetz thimbles is a valid description for any $k_0$.

The residues of finite $\Lambda$ poles of the einbein action vanish at the location of sources or ghost sources. 
The latter occur when oppositely oriented Lefschetz thimbles end at a single pole,  yielding canceling contributions to the inhomogeneous term in the Helmholtz equation.  Although ghost sources have no manifest physical interpretation, they are related to both cusp caustics and monodromies. 
The examples considered here suggest that cusp caustics lie on the domain of support of ghost sources, although this remains a conjecture.  The map between the einbein action and the catastrophe generating function is singular at the location of a ghost source. An improved understanding of this map is desirable to test the conjecture. Moreover, the possibility of generalizing the Thom-Arnold classification of catastrophes to the einbein action is tantalizing. 
The exactly soluble cases have meromorphic einbein actions and seem to only be capable of capturing $A_{N\le 3}$ caustics. It is conceivable that higher order caustics require einbein actions with singularities other than poles. 
Even if this is the case, the einbein action for an analytic index of refraction seems to only be capable of generating $A_N$ caustics.  Two einbein are apparently required to obtain the $D_N$, $E_6$,$E_7$, and $E_8$ catastrophes, for which the generating function is a polynomial in two variables.  Perhaps a two einbein description arises 
for a scatted field built from pairs of Green's functions coupled by a scattering kernel.


Much can be said about the analytic structure of the solution simply from the singularities of $\Psi$.
Monodromies under analytic continuation in the space of parameters defining the index of refraction, or in the spatial arguments of the Green's function, relate linearly independent solutions as well as different eigenrays.  These monodromies originate from the fact that the direction with which Lefschetz thimbles can approach poles of the einbein action varies with the argument of these parameters.
Interestingly, monodromies and cusp caustics seem to be two sides of the same coin, sharing an origin in the poles of $\mathbb S$. 

Remarkably, $\mathbb S$ has a much simpler dependence on $k_0$ than the full field obtained by integrating over $\Lambda$.  
At any given order in the Laurent series expansion of the einbein action about a pole,  the $1/k_0$ expansion truncates.  The constraints imposed by the Schr\"oedinger equation satisfied by $\Psi$ \eqref{Schrod}  suggest that a generalized Pad\'e approximant derived from the Laurent series about a pole of $\bar S$ may be highly accurate, perhaps even convergent, capturing the existence of other poles.  An explicit case is yet to be considered.


It is hoped that the einbein formulation may be of practical use in problems, especially those at low frequency or small $k_0$,  where regions of the shadow zone away from the caustic are of interest. In addition to missing arrivals due to complex rays, conventional approaches based on real ray theory only yield results in the immediate neighborhood of a caustic and are blind to the effects of perturbations in the shadow.  

The phenomenon of ray chaos may be interesting to consider within the framework of the einbein action.  
For an index of refraction giving rise to ray chaos,  the number of real eigenrays increases exponentially with range.  However, new poles of the einbein action are not generated by variation of $\vec x$.  The appearance of new real rays with range arises by passing through successive caustics, such that existing complex critical points merge and become real. 
The problem is then to construct an einbein action such that the number of caustics crossed with increasing range grows exponentially with range.

\section{Acknowledgements}

I thank Charles Spofford and Katherine Woolfe for multiple enlightening conversations on the subject of caustics in ocean acoustics.

\begin{appendices}

 \newpage
\section{Code displaying behavior of critical points upon crossing a cusp caustic}\label{appendixB}

\begin{figure}[!h]
	\center{
		\includegraphics[width= 100pt]{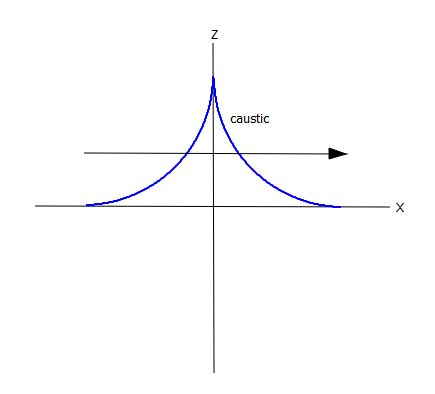}
		\caption{Crossing the cusp caustic.  }
		\label{fig:cuspcrossing}
	}
\end{figure}

\begin{figure}[!h]
	\center{
		\includegraphics[width= 300pt]{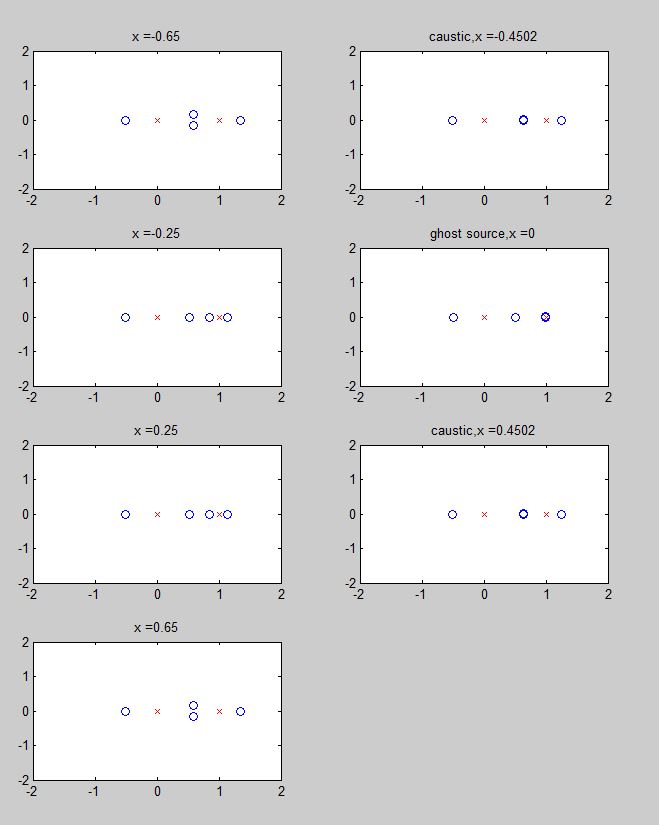}
		\caption{Snapshots of a crossing of the cusp caustic of section \ref{cusp} with $\mu=1,n_0=1,z=1$,  showing critical points and poles $\Lambda=0,1$ of the einbein action in the complex $\Lambda$ plane.  Poles are indicated by 'X'.  The pole $\Lambda=1$ is a ghost pole.  The negative critical point in the einbein description is  a spectator in this process,  having no Lefschetz thimble passing through it, and is not visible in the  uniform asymptotic description.}
		\label{fig:CrossingEinbein}
	}
\end{figure}

\begin{figure}[!h]
	\center{
		\includegraphics[width= 300pt]{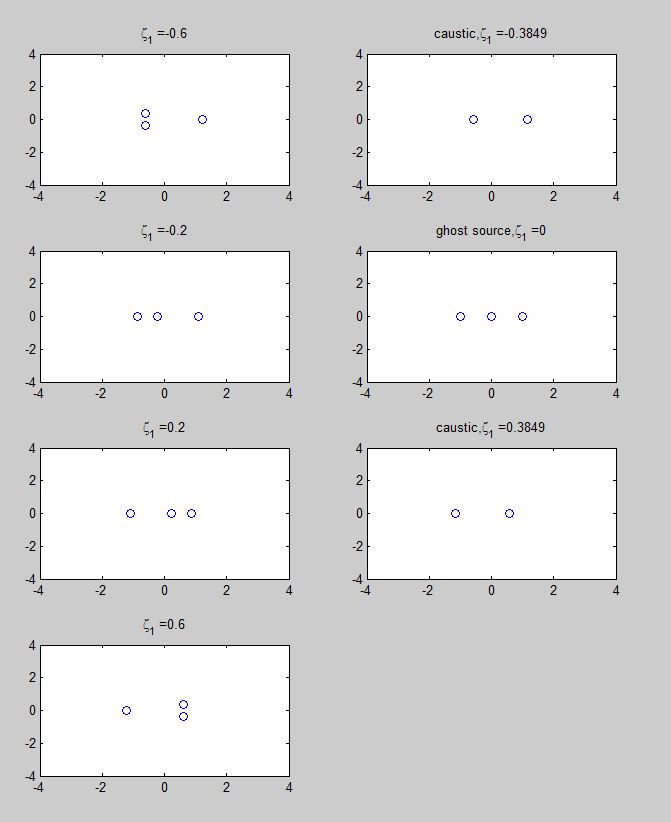}
		\caption{Snapshots of the crossing of the cusp caustic, showing the behavior of critical points of the generating function in the complex $\lambda$ plane. Each panel corresponds to the panel in the einbein description shown in figure \ref{fig:CrossingEinbein}.  In the complex $\lambda$ plane, there is no special behavior upon crossing the curve which maps to location of the ghost source, $\zeta_1=0$.}
		\label{fig:CrossingGenFunc}
	}
\end{figure}

It is illuminating to view an animation of the behavior of the critical points upon crossing through a cusp caustic, as shown in figure \ref{fig:cuspcrossing}.  The following code shows this process for the cusp described in section \ref{cusp}.  Results are generated in both the complex $\Lambda$ plane of the einbein description and in the complex $\lambda$ plane of the generating function, or uniform asymptotic, description.  For the einbein description, the poles of the einbein action are also plotted.  Snapshots of the output of this code at various points in the crossing are shown in figures \ref{fig:CrossingEinbein} and \ref{fig:CrossingGenFunc}.

\begin{verbatim}
%%% Motion of critical points in einbein description
z=1; %Change to z = 2 to intersect the singularity of the cusp caustic
xrange=[-2:0.01:2];
Nx=numel(xrange);
plot([0,1],[0,0],'rx'); %This plots the poles. the pole at [0,1] is a ghost pole.
for ind=1:Nx,
    x=xrange(ind);
    Poly=[ 4, -8, 4-x^2-z^2, 2*z^2, -z^2];
    Rts=roots(Poly); %These are the critical points. 
    repts = real(Rts);
    impts = imag(Rts);
    plot([0,1],[0,0],'rx');hold on;
    plot(repts,impts,'bo');
    xlim([-2 2]);
    ylim([-2,2]);
    pause(0.03);
    hold off;
end

%%% Motion of critical points in the generating function  description
zeta_2 = -1 %Change to zeta_2=0 to intersect the singularity of the cusp caustic
zeta_1_range=[-4:0.01:4];
Nzeta1=numel(zeta_1_range);
for ind=1:Nzeta1,    
    zeta_1=zeta_1_range(ind);   
    Poly=[1,0, zeta_2, zeta_1];
    Rts=roots(Poly);        
    repts = real(Rts);
    impts = imag(Rts);    
    plot(repts,impts,'bo');
    xlim([-4 4]);
    ylim([-4,4]);
    pause(0.005);    
end
\end{verbatim}



\newpage

\end{appendices}

\end{document}